\documentclass[preprintnumbers,superscriptaddress,showkeys,showpacs,byrevtex]{revtex4}
\usepackage{amsmath,amsfonts,amssymb,amscd,amsxtra,amsthm}
\usepackage{graphicx}
\usepackage{array} 
\usepackage{epstopdf}
\usepackage{bm}
\begin{document}   
\preprint{PKNU-NuHaTh-2015-02}
\title{Photoproduction of $\Lambda(1405)$ with the two-pole structure}
\author{Seung-il Nam}
\email[E-mail: ]{sinam@pknu.ac.kr}
\affiliation{Department of Physics, Pukyong National University (PKNU), Busan 608-737, 
Republic of Korea}
\affiliation{Asia Pacific Center for Theoretical Physics (APCTP), Pohang 790-784, 
Republic of Korea}
\author{Hyun-Kyu Jo}
\affiliation{Department of Physics, Pukyong National University (PKNU), Busan 608-737, Republic of Korea}
\date{\today}
\begin{abstract}
We investigate the $\Lambda(1405,1/2^-)\equiv\Lambda^*$ photoproduction off the proton target, i.e. $\gamma p\to K^+\Lambda^*$, considering explicitly its two-pole structure, the higher ($\Lambda^*_H:\,1430\,\mathrm{MeV}$) and lower ($\Lambda^*_L:\,1390\,\mathrm{MeV}$) mass-pole contributions, suggested by the chiral-unitary model (ChUM) approaches. For this purpose, we construct a two-body process model, which mimics the Dalitz process, $\gamma p\to K^+\pi\Sigma$, assuming that the mass of $\Lambda^*$ as the invariant mass of $\pi$ and $\Sigma$, i.e. $M_{\Lambda^*}\sim M_{\pi\Sigma}$. We employ the effective Lagrangian method with the tree-level Born approximation, using the gauge-invariant prescription for the phenomenological form factors. We provide the numerical results for the energy and angular dependences, $\pi$-$\Sigma$ invariant-mass distribution, and so on. It turns out that the model parameters determined from ChUM reproduce the experimental data qualitatively well, supporting the two-pole structure. Moreover, the nucleon resonance contribution near the threshold plays an important role to describe the data. 
\end{abstract} 
\maketitle
\section{Introduction}
$\Lambda(1405,1/2^-)\equiv\Lambda^*$ is the first excited state of the $\Lambda$-hyperon resonances. Among the peculiar and interesting features of this hyperon resonance to be addressed, its microscopic internal structure has been the most important issue for a couple of  decades: Which is the most genuine (or dominant) configuration for the $\Lambda^*$ internal structure within {\it one}- and {\it two-pole} ones? In conventional quark models, it had been taken into account as the $uds$ three-quark state, corresponding to the one-pole configuration, although its mass spectrum is hardly reproduced by the models: $M_{\Lambda^*}\approx1600$ MeV~\cite{Isgur:1978xj}. Even from the lattice QCD simulations, it was reported that the three-quark state for $\Lambda^*$ seems to be excluded, resulting in the higher mass $(1.7\sim1.8)$ MeV in various flavor-multiplet configurations~\cite{Nemoto:2003ft}.  The  mixture of the one- and two-pole contributions was suggested by the hybrid quark model~\cite{Nakamoto:2006br}. A similar scenario was also investigated via a few-body calculation using the isospin mixing, showing the two configurations are equivalently possible~\cite{Revai:2008wj}. In contrast, within some effective models, the one-pole structure for $\Lambda^*$ was preferred~\cite{Akaishi:2010wt,Nakamura:2013boa}. 

Being together with the chiral dynamics at low energy and the unitarity of scattering amplitudes, the meson-baryon interactions provide dynamical generations of baryon resonances in the coupled-channel methods: Chiral-unitary model (ChUM)~\cite{Magas:2005vu,Jido:2003cb,Jido:2002zk,Hyodo:2011ur}, providing a affirmative result for the $\Lambda^*$ mass. In this approach, $\Lambda^*$ shows a very interesting feature that the physically observed mass distribution for $\Lambda^*$ along the scattering line is a resultant interference between the higher- ($\sim1430$ MeV) and lower-mass ($\sim1390$ MeV) poles in the 2nd Riemann sheet~\cite{Hyodo:2011ur}. Although this is an interesting theoretical observation for the internal structure of the hyperon resonance, the two-pole structure scenario has not been proved obviously by experimental data so far. 

In the present work, we would like to study the photoproduction of $\Lambda^*$ off the proton target $\gamma p\to K^+\Lambda^*$, employing the effective Lagrangian method at the tree-level Born approximation, using the gauge-invariant prescription for the phenomenological form factors. Note that there have been several effective approaches for the photoproduction:  A simple Born approximation at tree-level calculation considering the $s$-channel dominance was studied in Ref.~\cite{Nam:2008jy}, taking account of the experimental data of the LEPS collaboration at Spring-8~\cite{Niiyama:2008rt}, but the theory failed to reproduce the recent CLAS data~\cite{Moriya:2013hwg}. In the crossing and duality consistent study,  it turned out that the total cross section is estimated as $\sigma\sim1\,\mu b$ for $\gamma p\to K^+\Lambda^*$, whereas $\sigma\sim1\,nb$ for the suppressed $\gamma K^-\to\gamma\Lambda^*$ process, due to the parity conservation~\cite{Williams:1991tw}. In Ref.~\cite{Nacher:1998mi}, the authors scrutinized the Dalitz process $\gamma p\to K^+\Lambda^*\to K^+\pi\Sigma$ by gauging the Weinberg-Tomozawa (WT) meson-baryon chiral interaction, giving the line shapes for the invariant-mass distribution for each isospin channel of $\pi\Sigma$. The Dalitz process was decomposed into the $K^- p\to \pi\Sigma$ process and the photon-kaon vertex, using the chiral coupled-channel approach information~\cite{Lutz:2004sg}. 

In the present work, taking the theoretical results of ChUM into account~\cite{Nam:2003ch}, we want to develop a simple model for a two-body scattering process which mimics the Dalitz process $\gamma p\to  K^+\pi\Sigma$. We assume that the higher- and lower-mass hypothetical $\Lambda^*$s, assigned by $\Lambda^*_H$ and $\Lambda^*_L$, couple to the physically measurable $\Lambda^*$, which can be understood as a $\pi$-$\Sigma$ quasi-bound state, so that its mass can be given as the $\pi$-$\Sigma$ invariant mass: $(k_\pi+k_\Sigma)^2=M^2_{\pi\Sigma}=M^2_{\Lambda^*}$. We compute the coupling strengths for $g_{\Lambda^*_{H,L}\Lambda^*}$, using the meson-baryon loop  diagram with the on-shell factorization and dimensional regularization. All the model parameters are taken from the ChUM calculations and experimental data. Especially, the parameters for $\Lambda^*_{H,L}$ are solely from ChUM, and we modify them to reproduce the recent CLAS data~\cite{Moriya:2013hwg}. The nucleon and hyperon resonance contributions, $N(2080)$ and $\Lambda(1670)$, are also taken into account. 

We provide the numerical results for the energy and angular dependences, $\pi$-$\Sigma$ invariant-mass distribution, $t$-channel momentum transfer, and photon-beam asymmetry. From the differential cross section $d\sigma_{\gamma p\to  K^+\Lambda^*}/d\cos\theta\equiv d\sigma/d\cos\theta$, in which $\theta$ indicates the outgoing kaon angle in the center-of-mass (cm) energy frame, we see that the CLAS experimental data~\cite{Moriya:2013hwg} are reproduced qualitatively well, whereas obvious underestimations are shown in the backward-scattering region, due the absence of possible $u$-channel contributions. It also turns out that the $K^*$-exchange contribution plays important role to describe the data, giving the coupling strength $g_{K^*N\Lambda^*}\approx-2.5$. Using the SU(6) relativistic quark-model calculations~\cite{Capstick:1998uh}, we consider the most dominant nucleon-resonance contribution, i.e. $N^*(2080)$, which dominates the region near the threshold with $(M,\Gamma)_{N(2080)}\approx(2000,230)$ MeV, while the $\Lambda(1670)$ contribution plays only minor role with the ChUM information. In reproducing the data, we choose the full decay widths for $\Lambda^*_{H,L}$ as $\Gamma_{H,L}=(30,70)$ MeV, which deviate from the ChUM estimations $\Gamma_{H,L}=(14,74)$ MeV. Moreover, the phase angle between the invariant amplitudes becomes $\phi=2.83$, which is about $90\%$ of the ChUM estimation. The reason for these deviations can be understood by that we have different (or small) background contributions in the present model calculations in comparison to the ChUM coupled-channel ones. Once all the model parameters are fixed for $d\sigma/d\cos\theta$, the total cross section as a function of $E_\gamma$ shows good agreement with the experimental data, manifesting the dominant $N^*$ contribution near the threshold and sizable contribution from the $K^*$ exchange over the broad energy range. 

Since the $\Lambda^*$ mass is identified with $M_{\pi\Sigma}$ in our model, and $\Lambda^*$ couples to intermediate $\Lambda^*_{H,L}$, we can analyze the cross section as a function of the invariant mass. Assuming that the full-decay widths for the intermediate states are small enough in comparison to their mass, the invariant-mass distribution for the Dalitz process $\gamma p\to  K^+\pi\Sigma$ can be rewritten with the two-body process. By doing this, we compute $d\sigma_{\gamma p\to  K^+\pi\Sigma}/dM_{\pi\Sigma}$ as a function of $W$, which denotes the cm energy $E_\mathrm{cm}=\sqrt{s}$, and the numerical results provide qualitative agreement with the experimental data~\cite{Moriya:2013eb}. As for the region $M_{\pi\Sigma}=(1.355\sim1.455)$ GeV, the mass distribution is relatively symmetric about $M_{\pi\Sigma}\approx1.4$ GeV for the lower $W$, due to the strong $N^*$ contribution. As the energy increases, the mass distribution becomes slightly asymmetric, since the two pole contributions interfere each other and dominate the region with the diminishing $N^*$ contribution. This observation is confirmed once again via the invariant-mass distribution as a function of $M_{\pi\Sigma}$ and $E_\gamma$ comparing with the data.  The $t$-channel momentum-transfer $d\sigma/dt$ shows quiet typical behavior as shown in other photoproduction processes. The photon-beam asymmetry $\Sigma$ indicates the strong $K$-exchange contribution in the $t$ channel, and the $K^*$-exchange one becomes manifest as the energy increases, although $\Sigma$ is all positive for the energy range $W=(2.0\sim2.8)$ GeV. 

From all the observations discussed above, we can conclude that the present model calculations, which manifest the two-pole structure for $\Lambda(1405)$, reproduce the presently available experimental data qualitatively well with the help of the theoretical (mostly from ChUM) and experimental information. This leads us to the consequence that the two-pole structure scenario seems quite supporting. However, we also accept that the similar consequence can be acquired by a single-pole scenario, i.e. $\Lambda^*_{H,L}\to\Lambda^*_\mathrm{single}$, and it looks quite difficult still to pin down the genuine internal structure of $\Lambda^*$ from the present effective approach. 

The present work is organized as follows: Section II will be devoted to explain how to construct the effective Lagrangian model, which mimics a Dalitz process, considering the two-pole structure. The numerical results and relevant discussions are given in Section III, and the summary, conclusion, and future perspectives in Section IV.
\section{Theoretical framework}
\begin{figure}[h]
\includegraphics[width=16cm]{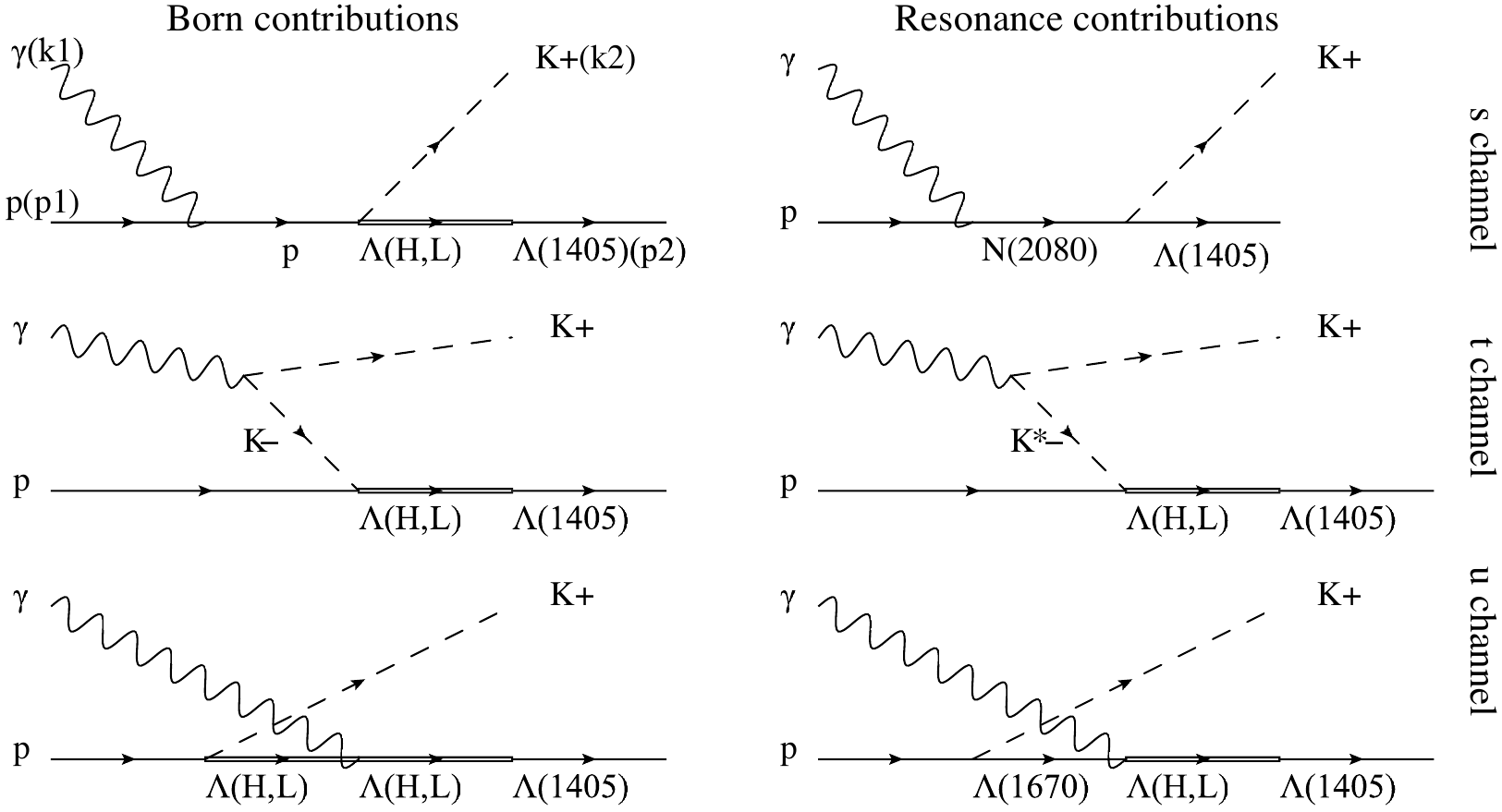}
\caption{Tree-level Feynman diagrams for the $\gamma p\to K^+\Lambda(1405)$ reaction process with the two-pole structure of $\Lambda(1405)$ in the present work, considering the Born (left column) and resonance (right column) contributions in the $(s,t,u)$ channels. The higher- and lower-pole contributions of $\Lambda(1405)$ are indicated by $\Lambda(H:1430\,\mathrm{MeV})$ and $\Lambda(L:1390\,\mathrm{MeV})$, respectively, here. The four momenta for the incident photon $(\gamma)$, target proton $(p)$, outgoing kaon $(K^+)$, and recoil $\Lambda^*$ are denoted by $k_1$, $p_1$, $k_2$, and $p_2$, respectively.}       
\label{FIG1}
\vspace{1cm}
\includegraphics[width=16cm]{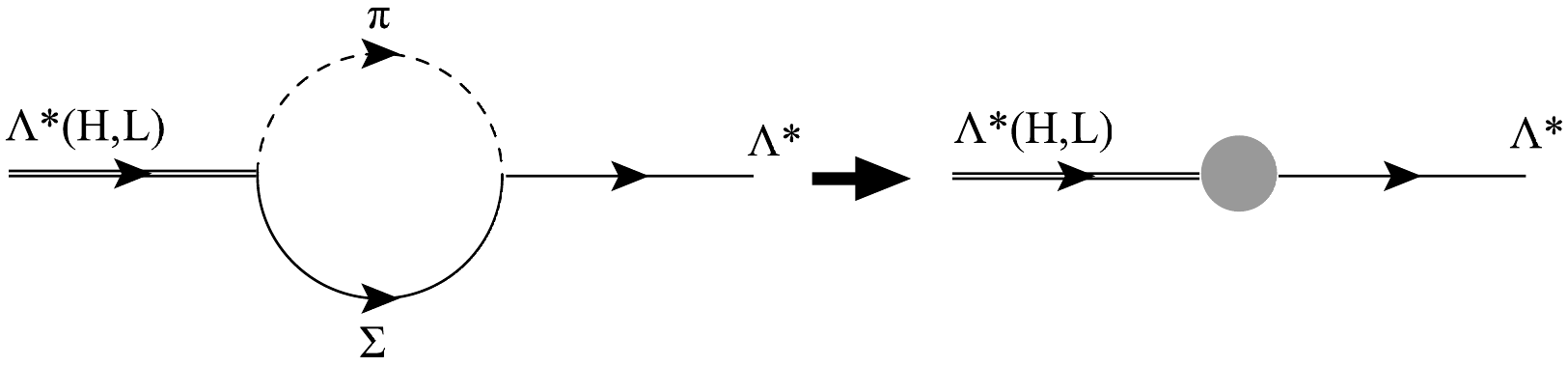}
\caption{In the left, $\Lambda^*_{H,L}$ couple to $\Lambda^*$ via the $\pi$-$\Sigma$ loop, considering that the physical $\Lambda(1405)$ state decays almost into the $\pi$-$\Sigma$ channel $\sim100\%$. In the right, we show the effective baryon-baryon vertex, deduced from the $\pi$-$\Sigma$ loop diagram.}       
\label{FIG3}
\end{figure}

We would like to start explaining theoretical framework for investigating the $\gamma p\to K^+\Lambda^*$ reaction process with the two-pole structure. As shown in Fig.~\ref{FIG1}, we consider the six Feynman diagrams in total. We assign the four momenta for the incident photon, target proton, outgoing kaon, and recoil $\Lambda^*$ as $k_1$, $p_1$, $k_2$, and $p_2$, respectively. Our main assumption is that the physically identified $\Lambda^*$ resonance, reconstructed from the decaying $\pi$ and $\Sigma$ in the Dalitz process $\gamma p\to K^+\Lambda_{H,L}\to K^+\pi\Sigma$, can be  understood as a {\it physical} $\Lambda^*$. Hence, it can be thought that the mass of $\Lambda^*$ corresponds to the invariant mass $M_{\pi\Sigma}\equiv M(\pi\Sigma)=M_{\Lambda^*}$. The two {\it hypothetical} states, which were suggested theoretically by ChUM, are assigned by $\Lambda^*_H$ for the higher mass state ($\sim1430$ MeV) and the lower one $\Lambda^*_L$ ($\sim1390$ MeV).  Therefore, we construct interaction vertices in which $\Lambda^*$ is coupled to $\Lambda^*_H$ as well as $\Lambda^*_L$, resembling the Dalitz process, at the amplitude level, as explicitly demonstrated in Fig.~\ref{FIG1}. In the present work, we consider the $s$-, $t$-, and $u$-channel contributions. For the $s$ channel, we consider the ground-state and resonance nucleons. The pseudoscalar- and vector-kaon exchanges, $K$ and $K^*$, are taken into account for the $t$ channel, whereas the $\Lambda^*_{H,L}$ and hyperon resonance contributions for the $u$ channel. Take notice of that we do not consider the possibility that the $p$-wave $\Sigma(1385,3/2^+)$ contribution, which can couple to $\Lambda^*$ for brevity in the present work. 

For computing the invariant amplitudes for the Feynman diagrams given in Fig.~\ref{FIG1}, we define the effective Lagrangians for the interaction vertices as follows:
\begin{eqnarray}
\label{eq:EFFLAG}
\mathcal{L}_{\gamma KK}
&=&ie_K\left[
(\partial^{\mu}K^{\dagger})K-(\partial^{\mu}K)K^{\dagger}
\right]A_{\mu}+ {\rm h.c.},\,\,\,\,
\mathcal{L}_{\gamma{NN}}=
-\bar{N}\left[e_N\rlap{\,/}{A}
+\frac{e\kappa_{N}}{2M_N}
\sigma_{\mu\nu}F^{\mu\nu}\right]N+ {\rm h.c.},
\cr
\mathcal{L}_{\gamma\Lambda^*\Lambda^*}
&=&
-\frac{e\kappa_{\Lambda^*}}{2M_{\Lambda^*}}
\bar{\Lambda}^*\sigma_{\mu\nu}F^{\mu\nu}\Lambda^*+ {\rm h.c.},
\,\,\,\,
\mathcal{L}_{\gamma K K^*}= g_{\gamma K K^{*}}
\epsilon_{\mu\nu\sigma\rho}(\partial^{\mu}A^{\nu})
(\partial^{\sigma}K^{\dagger}){K}^{*\rho}, 
\cr
\mathcal{L}_{K^{*}N\Lambda^*_i}&=&g_{K^{*}N\Lambda^*_i}\bar{\Lambda}^*_i
\gamma^{\mu}\gamma_{5}{K}^{*\dagger}_{\mu}N
+ {\rm h.c.}, 
\,\,\,\,
\mathcal{L}_{KN\Lambda^*_i}=ig_{KN\Lambda^*_i}
\bar{\Lambda}^*_iKN,\,\,\,\,\mathcal{L}_{\pi\Sigma\Lambda^*_i}=ig_{\pi\Sigma\Lambda^*_i}
\bar{\Lambda}^*_i\bm{\pi}\cdot\bm{\Sigma},
\cr
\mathcal{L}_{\pi\Sigma\Lambda^*}&=&ig_{\pi\Sigma\Lambda^*}
\bar{\Lambda}^*\bm{\pi}\cdot\bm{\Sigma},\,\,\,\,
\mathcal{L}_{\Lambda^*\Lambda^*_i}=g_{\Lambda^*\Lambda^*_i}\bar{\Lambda}^*\Lambda^*_i,
\end{eqnarray}
where $K$, $K^*_\mu$, $A_\mu$, $N$, $\Lambda^*$, and $\Lambda^*_i$, and  denote the fields for the pseudo, vector kaons, photon, nucleon, $\Lambda(1405)$, and the hypothetical $\Lambda^*$ contributions for higher and lower mass ones, i.e. $i=(H,L)$. We also consider $\pi$ and $\Sigma$ fields, coupling to $\Lambda^*_i$ for the later use. We set $e_h$, $M_h$, and $\kappa_h$ for the electric charge, mass, and anomalous magnetic moment for the hadron $h$, respectively, whereas $e$ stands for the unit electric charge. For instance, we have $e_p=(+1)e$. As for the relevant strong coupling constants are given by $g_{h_1h_2h_3}$, their values are determined by experimental and theoretical information, listed in Table~\ref{TAB1}. Note that all the couplings for $\Lambda^*_{H,L}$ are taken from the ChUM coupled-channel calculations. 
\begin{table}[b]
\begin{tabular}{>{\centering}m{2cm}|
>{\centering}m{2cm}|
>{\centering}m{2cm}|
>{\centering}m{2cm}|
>{\centering}m{2cm}|
>{\centering\arraybackslash}m{2cm}}
$\kappa_p$&$\kappa_{\Lambda^*_{H,L}}$&$g_{\gamma K^+ K^{*-}}$&$g_{KN\Lambda^*_{H,L}}$&$g_{\pi\Sigma\Lambda^*_{H,L}}$&$g_{\pi\Sigma\Lambda^*}$\\
\hline
$1.79$
&$0.41,0.30$~\cite{Jido:2002yz}
&$0.254/\mathrm{GeV}$
&$2.52,1.43$~\cite{Nam:2003ch}
&$1.30,2.06$~\cite{Nam:2003ch}
&$0.91$~\cite{Agashe:2014kda}\\
\end{tabular}
\caption{Relevant EM and strong coupling constants.}
\label{TAB1}
\end{table}

In order to determine  $g_{\pi \Sigma\Lambda^*}$, we have used the following equation with the experimental data~\cite{Agashe:2014kda}:
\begin{equation}
\label{eq:GAMMA}
\Gamma_{\Lambda^*\to\pi\Sigma}=
\frac{3g^2_{\pi\Sigma\Lambda^*}|\bm{k}|(M_\Sigma+E_\Sigma)}{4\pi M_{\Lambda^*}}
\approx50\,\mathrm{MeV},
\end{equation}
where 
\begin{equation}
\label{eq:THREE}
|\bm{k}|=|\bm{k}_\Sigma|=|\bm{k}_\pi|=\frac{\sqrt{[M^2_{\Lambda^*}-(M_\Sigma-M_\pi)^2]
[M^2_{\Lambda^*}-(M_\Sigma+M_\pi)^2]}}{2M_{\Lambda^*}}\approx369\,\mathrm{MeV},
\,\,\,\,
E_\Sigma=\sqrt{M^2_\Sigma+\bm{k}^2_\Sigma}.
\end{equation}
Since the information to determine $g_{K^{*}N\Lambda^*_i}$ is poor, we consider it as a free parameter of the present model. Using the effective Lagrangians given in Eq.~(\ref{eq:EFFLAG}), it is straightforward to compute the invariant amplitudes as follows:
\begin{eqnarray}
\label{eq:BORNAMP}
i\mathcal{M}^s_i&=&e^{i\phi_i}
g_{\Lambda^*_i\Lambda^*}g_{KN\Lambda^*_i}
\bar{u}_{\Lambda}(p_2)\left[\frac{\rlap{/}{p}_2+M_i}{M^2_{\pi\Sigma}-M^2_i+i\Gamma_iM_i} \right]\left[\frac{\rlap{/}{q}_s+M_p}{s-M^2_p} \right]
\left[e_p\rlap{/}{\epsilon}+\frac{e\kappa_p}{2M_p}\rlap{/}{k}_1\rlap{/}{\epsilon} \right]
u_p(p_1),
\cr
i\mathcal{M}^{t_K}_i&=&2e^{i\phi_i}
g_{\Lambda^*_i\Lambda^*}g_{KN\Lambda^*_i}
\bar{u}_{\Lambda}(p_2)
\left[\frac{\rlap{/}{p}_2+M_i}{M^2_{\pi\Sigma}-M^2_i+i\Gamma_iM_i} \right]
\left[\frac{(k_2\cdot\epsilon)}{t-M^2_K} \right]u_p(p_1),
\cr
i\mathcal{M}^{t_{K^*}}_i&=&ig_{\gamma KK^*}e^{i\phi_i}
g_{\Lambda^*_i\Lambda^*}g_{K^*N\Lambda^*_i}
\bar{u}_{\Lambda}(p_2)\left[\frac{\rlap{/}{p}_2+M_i}{M^2_{\pi\Sigma}-M^2_i+i\Gamma_iM_i} \right]
\left[\frac{\gamma_5(\epsilon_{\mu\nu\sigma\rho}
k^\mu_1\epsilon^\nu k^\sigma_2\gamma^\rho)}
{t-M^2_{K^*}+i\Gamma_{K^*}M_{K^*}} \right]u_p(p_1), 
\cr
i\mathcal{M}^u_i&=&e^{i\phi_i}\frac{e\kappa_{\Lambda^*_i}}{2M_{\Lambda^*_i}}
g_{\Lambda^*_i\Lambda^*}g_{KN\Lambda^*_i}
\bar{u}_{\Lambda}(p_2)\left[\frac{\rlap{/}{p}_2+M_i}{M^2_{\pi\Sigma}-M^2_i+i\Gamma_iM_i} \right]
\rlap{/}{\epsilon} \rlap{/}{k}_1\left[\frac{\rlap{/}{q}_s+M_p}{u-M^2_{\Lambda^*_i}} \right]u_p(p_1),
\end{eqnarray}
where $\phi_i$ is a phase angle between the invariant amplitudes $i\mathcal{M}_{i=H,L}$. In the ChUM calculations, the value for $\phi$ is estimated to be about $\pi$~\cite{phaseangle}. It is worth mentioning that, since we consider $\Lambda^*$ as the $\pi$-$\Sigma$ bound state, its mass can be chosen as $M_{\pi\Sigma}$ as already discussed above. If this is the case, the four momentum for $\Lambda^*$ must satisfy the condition that $p^2_2=M^2_{\Lambda^*}=M(\pi\Sigma)\equiv M^2_{\pi\Sigma}$, shown in the propagator with $p_2$ in Eq.~(\ref{eq:BORNAMP}). By doing this, we will take $M_{\pi\Sigma}$ as a dynamical variable in the present work. In this setup, the four momentum for the particles can be defined explicitly in the cm frame as follows:
\begin{eqnarray}
\label{eq:MOMENTA}
k_\gamma&\equiv&k_1=\left(\bm{k}_i,0,0,\bm{k}_i\right),
\cr
p_p&\equiv&p_1=\left(\sqrt{\bm{k}^2_i+M_p},0,0,-\bm{k}_i\right),
\cr
k_{K^+}&\equiv&k_2=\left(\sqrt{\bm{k}^2_f+M_K},\bm{k}_f\sin\theta,0,\bm{k}_f
\cos\theta\right),
\cr
p_{\Lambda^*}&\equiv&p_2=
\left(\sqrt{\bm{k}^2_f+M_{\pi\Sigma}},-\bm{k}_f\sin\theta,0,-\bm{k}_f\cos\theta\right).
\end{eqnarray}
Here, $\bm{k}_{i,f}$ and $\theta$ indicate the three momenta for the initial and final states, and the angle for the outgoing kaon in the cm frame. Here, the reaction plane is defined by the $x$-$z$ plane, whereas the $y$ axis is perpendicular to the plane. Note that the second square brackets in the r.h.s. of Eq.~(\ref{eq:BORNAMP}) is a obviously new term in the present model which is not seen in usual tree-level Born-approximation approaches, and indicates a propagating intermediate particle, such as the $\Lambda^*_{H,L}$, giving a invariant-mass distribution with its maximum at $M_{\pi\Sigma}=M_i$ with the width $\Gamma_i$. In the ChUM calculations, the values for the full-decay widths for the hypothetical particles are given by 
$\Gamma_{H,L}\approx(14,74)$ MeV~\cite{Nam:2003ch}. It is clear that the sum of all the bare amplitude in Eq.~(\ref{eq:BORNAMP}) satisfies the Ward-Takahashi identity.

Now, we are in a position to determine the value for the coupling constant $g_{\Lambda^*_i\Lambda^*}$, which represents the coupling strength between the hypothetical and physical $\Lambda^*$s. We note that this coupling can be understood as a meson-baryon loop as shown in Fig.~\ref{FIG3}. Since the physical $\Lambda^*$ decays into $\pi$ and $\Sigma$ in about $100\%$ experimentally~\cite{Agashe:2014kda}, we only consider the $\pi$-$\Sigma$ loop here for simplicity. Hence, the relevant effective Lagrangian in Eq.~(\ref{eq:BORNAMP}), $\mathcal{L}_{\Lambda^*\Lambda^*_i}$ can be rewritten as 
\begin{equation}
\label{eq:EFFG}
\mathcal{L}_{\Lambda^*\Lambda^*_i}\to
g_{\Lambda^*\Lambda^*_i}(k^2)\,\bar{u}_{\Lambda^*}(k)
u_{\Lambda^*_i}(k)
\approx g_{\pi\Sigma\Lambda^*_i}g_{\pi\Sigma\Lambda^*_{P}}\bar{u}_{\Lambda^*}(k)
\left[i\int\frac{d^4q}{(2\pi)^4}\frac{2M_\Sigma}{[q^2-M^2_\Sigma][(k-q)^2-M^2_\pi]} \right]
u_{\Lambda^*_i}(k).
\end{equation}
In deriving Eq.~(\ref{eq:EFFG}), we make use of the {\it on-shell} factorization, which has been employed in the ChUM calculations widely, and it makes calculations easier to a good extent. The loop integral can be performed with the dimensional regularization~\cite{Nam:2003ch}, and we can obtain the relevant coupling strength $g_{\Lambda^*\Lambda^*_i}$ as a function of the invariant mass $M_{\pi\Sigma}$:
\begin{equation}
\label{eq:GVALUE}
g_{\Lambda^*\Lambda^*_i}(M^2_{\pi\Sigma})=g_{\pi\Sigma\Lambda^*_i}g_{\pi\Sigma\Lambda^*}
\frac{2M_\Sigma}{16\pi^2}
\left[\frac{M^2_\pi-M^2_\Sigma+M^2_{\pi\Sigma}}{2M^2_{\pi\Sigma}}\ln\frac{M^2_\pi}{M^2_\Sigma}
+\frac{\xi}{2M^2_{\pi\Sigma}}\ln\frac{M^2_\pi+M^2_\Sigma-M^2_{\pi\Sigma}-\xi}{M^2_\pi+M^2_\Sigma-M^2_{\pi\Sigma}+\xi}+\ln\frac{M^2_\Sigma}{\mu^2} \right],
\end{equation}
where $\xi=\sqrt{[M^2_{\pi\Sigma}-M_\Sigma-M_\pi)^2][M^2_{\pi\Sigma}-(M_\Sigma+M_\pi)^2]}$. To tame the UV divergence in the loop integral, we use the dimensional regularization, and the renormalization scale $\mu$ is chosen to be $2.0$ GeV, which was determined by fitting the data in the ChUM calculations for $S=-1$ channel~\cite{Hyodo:2011ur}. Numerically, we have $g_{\Lambda^*\Lambda^*_1}=-(41.59\sim39.25)$ MeV and $g_{\Lambda^*\Lambda^*_2}=-(26.24\sim24.76)$ MeV for $M_{\pi\Sigma}=(1.355\sim1.455)$ MeV. It is worth mentioning that it is also possible for the incident photon can couple to the $\pi$-$\Sigma$ loop in principle. By doing this, i.e. with the electromagnetic interactions with the hadrons, one can extract separate information for each isospin channel of $\pi$-$\Sigma$. However, we are interested in the isospin sum or average in the present work, we want leave this interesting issue for the future work.

The nucleon and hyperon resonance contributions can be of importance. As for the nucleon resonances above $2.0$ GeV, we have $D^{**}_{13}(2080)$, $S^*_{13}(2090)$, $P^*_{11}(2100)$, $G^{****}_{17}(2190)$, $D^{**}_{15}(2200)$, $H^{****}_{19}(2200)$, $G^{****}_{19}(2250)$, and so on~\cite{Nakamura:2010zzi}. In the recent PDG listing~\cite{Agashe:2014kda}, $D_{13}(2080)$ is now split into $D_{13}(2120)$ and  $D_{13}(1875)$ by the multi-channel partial-wave analyses. In the present work, to be consistent with theoretical information, i.e. the SU(6) quark model, as will be discussed below, we choose the nucleon-resonance masses and helicity amplitudes from Ref.~\cite{Nakamura:2010zzi}. In the relativistic SU(6) quark model, the strong partial-wave decay amplitude (PWDA) $G_{KN^*\Lambda^*}(\ell)$ is estimated for the nucleon resonances of Ref.~\cite{Capstick:1998uh}, being listed in  Table~\ref{TAB2}. By definition, the PWDA corresponds to the decay width as
\begin{equation}
\label{eq:PWDA}
\Gamma_{N^*\to K\Lambda^*}=\sum_\ell|G_{KN^*\Lambda^*}(\ell)|^2.
\end{equation}
Thus, using Eq.~(\ref{eq:PWDA}), one can obtain the relevant strong coupling constants, while $\Gamma_{N^*\to K\Lambda^*}$ in the l.h.s. of Eq.~(\ref{eq:PWDA}) can be computed by the relevant effective Lagrangians.

\begin{table}[b]
\begin{tabular}{c|ccccccc}
&$D^{**}_{13}(2080)$&$S^{*}_{13}(2090)$&$P^{*}_{11}(2100)$&$G^{****}_{17}(2190)$&$D^{**}_{15}(2200)$&$H^{****}_{19}(2200)$&$G^{****}_{19}(2250)$\\
\hline
$G_{KN^*\Lambda^*}(\ell)$&$3.9^{+1.3}_{-2.7}$&$0.5^{+1.0}_{-0.4}$&$5.2\pm0.8$&$1.2\pm0.7$&$0.0\pm0.0$&$-0.3^{+0.2}_{-0.3}$&$0.0\pm0.0$\\
\end{tabular}
\caption{Strong partial-wave decay amplitude $G_{KN^*\Lambda^*}(\ell)$~\cite{Capstick:1998uh}.}
\label{TAB2}
\end{table}

Ignoring the less-confirmed $(*)$ and relatively small-coupling resonances from Table~\ref{TAB2}, we find that $D_{13}$ is the most dominant contribution. Therefore, for simplicity, we only consider this resonance hereafter. The relevant effective Lagrangians for the $D_{13}$ resonance contribution as shown in Fig.~\ref{FIG1} read:
\begin{eqnarray}
\label{eq:NSTAR}
\mathcal{L}_{\gamma ND_{13}}&=&-ie\left[
\frac{h_1}{2M_N}\bar{N}\gamma_{\nu}
-\frac{ih_2}{(2M_N)^3}(\partial_\nu \bar{N})\gamma_5 \right]F^{\mu\nu}D_{13\mu},
\cr
\mathcal{L}_{KD_{13}\Lambda^*}&=&\frac{g_{KD_{13}\Lambda^*}}{M_K}\bar{D}_{13\mu}(\partial^{\mu}K)\Lambda^*.
\end{eqnarray}
Using the relativistic SU(6) quark-model information for PWDA and effective Lagrangian in Eq.~(\ref{eq:NSTAR}), numerically, we have $g_{KD_{13}\Lambda^*}\approx1.16$, considering only the dominant $s$-wave ($\ell=0$) contribution near the threshold. Similarly, employing the experimental data for the helicity amplitudes~\cite{Agashe:2014kda}, one obtains $(h_1,h_2)=(-0.83,+2.14)$. The invariant amplitude becomes
\begin{eqnarray}
\label{eq:RESO}
i\mathcal{M}^{D_{13}}&=&-\frac{eh_1g_{KD_{13}\Lambda^*}}{2M_NM_K}
\bar{u}_\Lambda(p_2)\frac{(\rlap{/}{q}_s+M_{D_{13}})(\mathcal{A}_{4\times4}-\mathcal{B}_{4\times4})}{s-M^2_{D_{13}}+iM_{D_{13}}\Gamma_{D_{13}}}u_N(p_1),
\cr
\mathcal{A}_{4\times4}&=&\left[k_1\cdot k_2-\frac{1}{3}\rlap{/}{k}_2\rlap{/}{k}_1-\frac{1}{3M_{D_{13}}}
[(k_1\cdot q_s)\rlap{/}{k}_2-(k_2\cdot q_s)\rlap{/}{k}_1]-\frac{2}{3M^2_{D_{13}}}(k_1\cdot q_s)(k_2\cdot q_s) \right]\rlap{/}{\epsilon},
\cr
\mathcal{B}_{4\times4}&=&\mathcal{A}_{4\times4}(k_1\leftrightarrow \epsilon).
\end{eqnarray}
Note that we set $h_2$ to be zero here, since its contribution in the relatively low-energy region must be small. As understood in Eq.~(\ref{eq:RESO}), we do not consider the intermediate $\Lambda^*_{H,L}$ for this contribution, i.e. the nucleon resonance couples directly to $\Lambda^*$. As for the hyperon resonance, we may take $\Lambda(1670,1/2^-)\equiv\Lambda^\star$ into account as shown in Fig.~\ref{FIG1}. However, the transition magnetic couplings computed from ChUM~\cite{Jido:2002yz} are too small to provide sizable contribution in comparison to others, $\kappa_{\gamma\Lambda^*_{H,L}\Lambda^{\star}}=(0.019\pm0.002,0.093\pm0.003)$ for instance, we drop those hyperon resonance contribution. We verified that these contributions are negligible indeed numerically. 

Since the hadrons are spatially extended objects, we need to take the phenomenological form factors into account. Following the gauge-invariant form-factor prescription suggested and employed in Refs.~\cite{Haberzettl:1998eq,Davidson:2001rk,Haberzettl:2006bn,Nam:2005uq}, we define the relevant form factors as follows:
\begin{equation}
\label{eq:FF}
F_c=F_{s,p}+F_{t,K}-F_{s,p}F_{t,K},\,\,\,\,
F_{x,h}=\frac{\Lambda^4_{h}}{\Lambda^4_{h}+(x-M^2_h)^2},
\end{equation}
where $x$ and $h$ denote the Mandelstam variable $x=(s,t,u)$ and hadron species $h$. Then, the dressed invariant total amplitude  for the present reaction process can be written in the gauge-invariant form factor scheme:
\begin{equation}
\label{eq:BORN}
i\mathcal{M}_\mathrm{total}=\sum_ie^{i\phi_i}\left[(i\mathcal{M}^s_i+i\mathcal{M}^{t_K}_i)F_c +i\mathcal{M}^u_iF_{u,\Lambda^*_i}+i\mathcal{M}^u_iF_{t,K^*}\right]+i\mathcal{M}^{D_{13}}F_{s,D_{13}}
\end{equation}
Note that, for brevity, we choose $\Lambda_h=1.0$ GeV in common for all the hadrons throughout this work. 

\section{Numerical results and Discussions}
\begin{figure}[t]
\begin{tabular}{ccc}
\includegraphics[width=6cm]{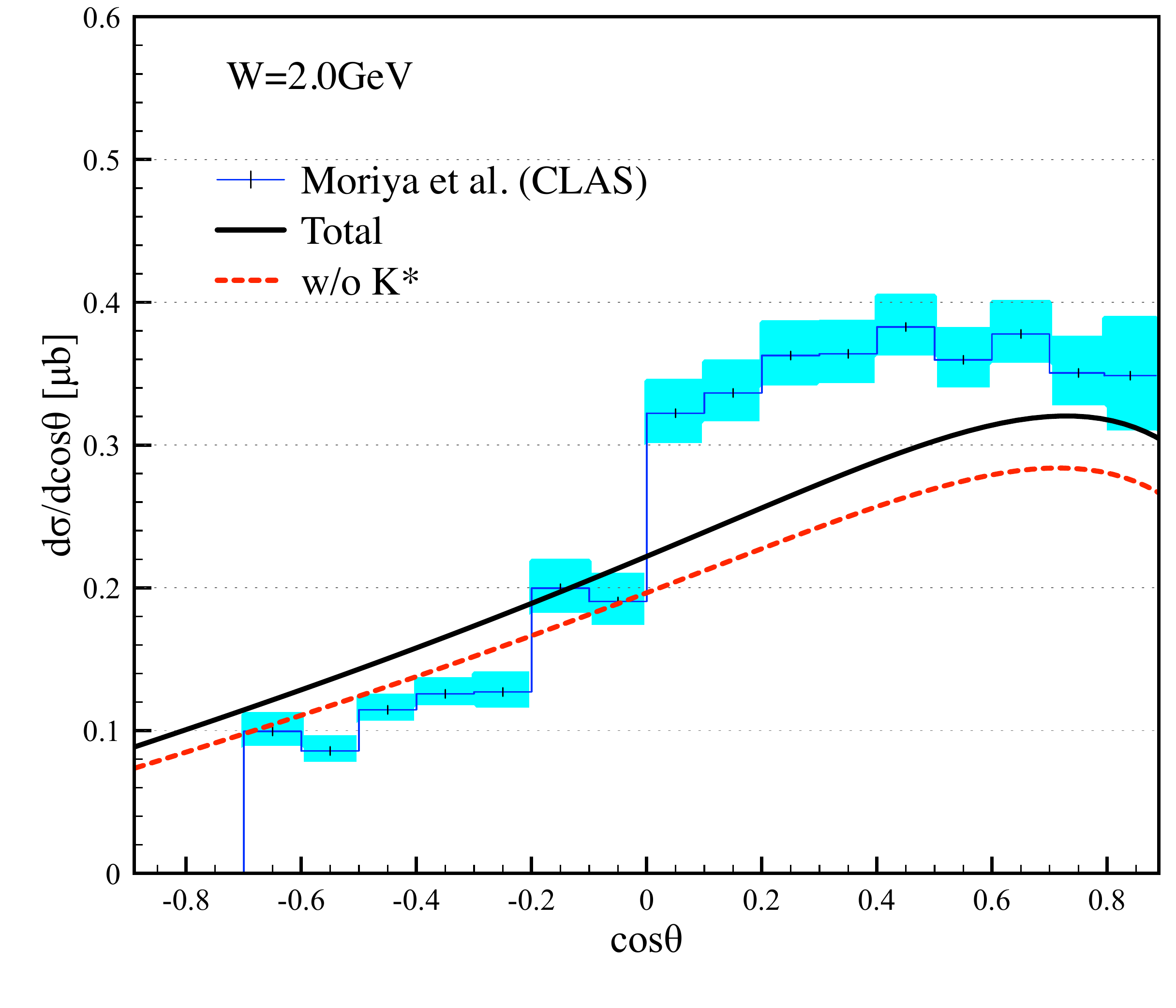}
\includegraphics[width=6cm]{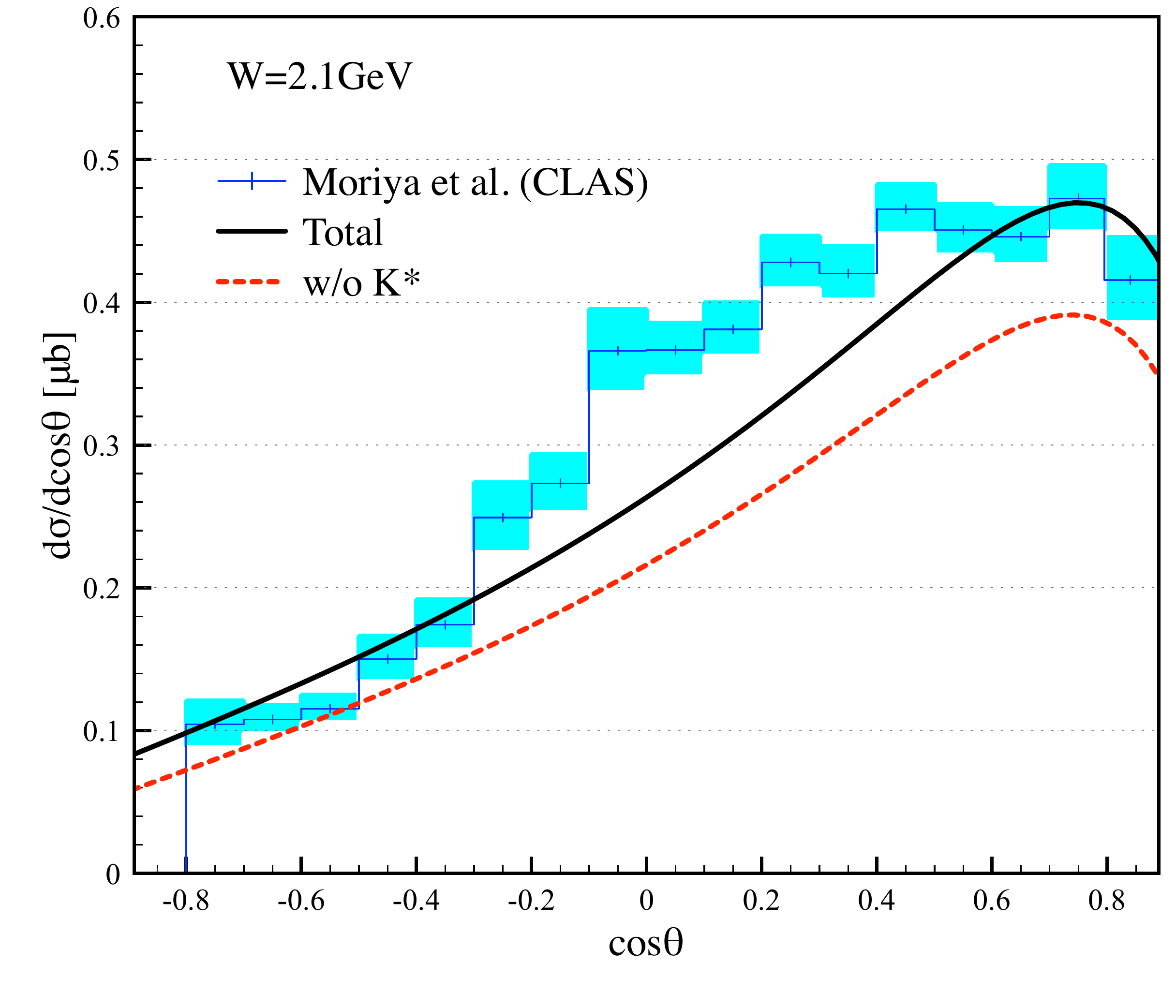}
\includegraphics[width=6cm]{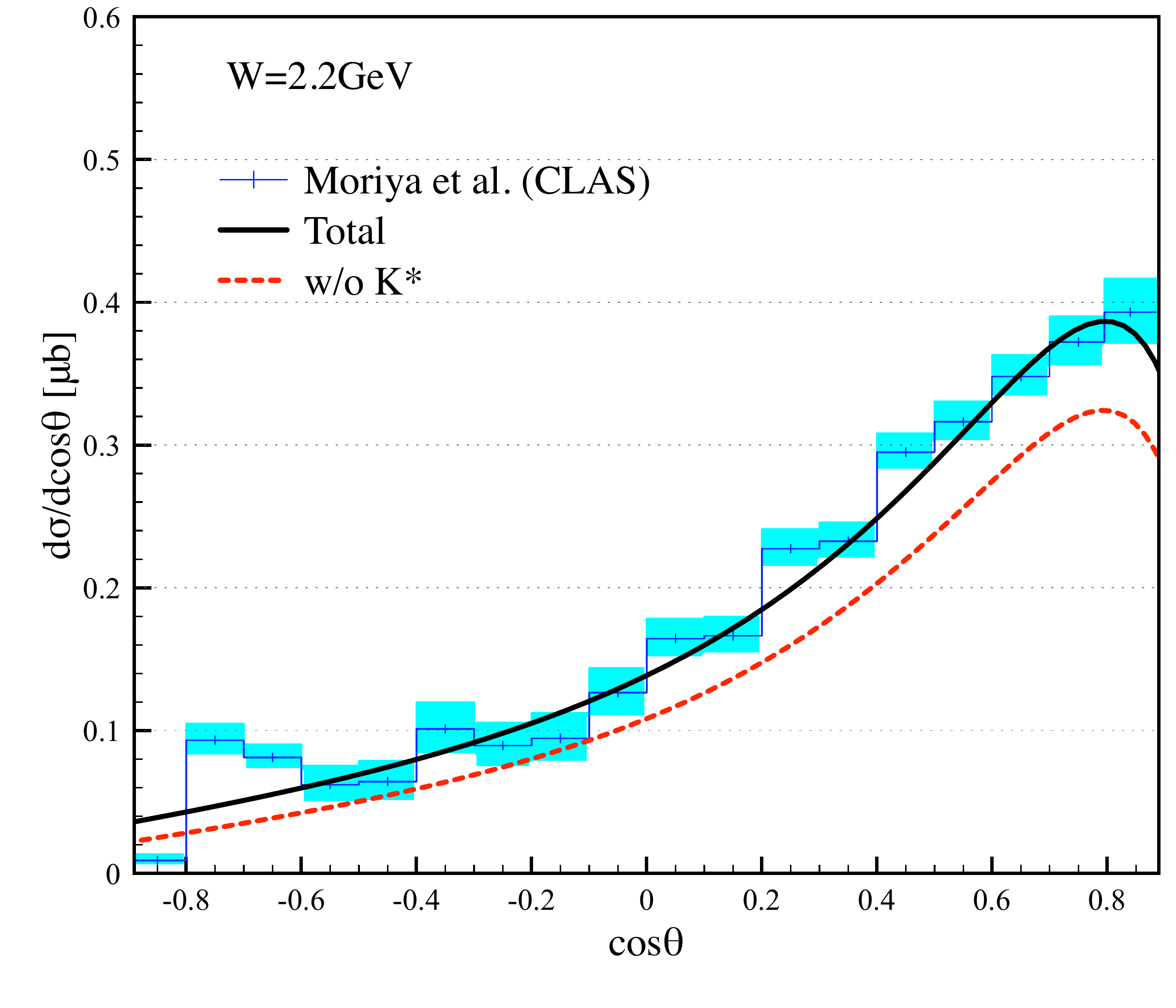}
\end{tabular}
\begin{tabular}{ccc}
\includegraphics[width=6cm]{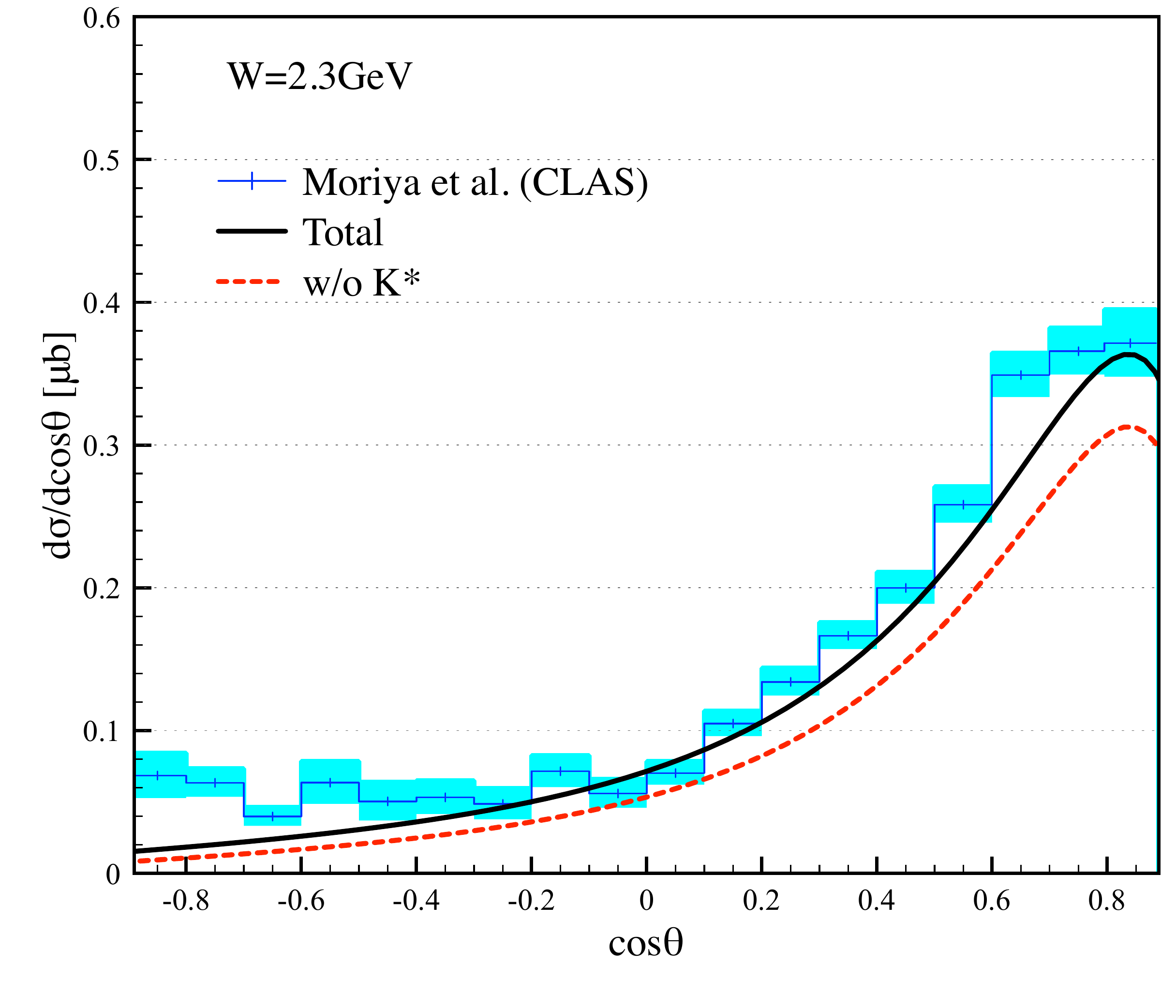}
\includegraphics[width=6cm]{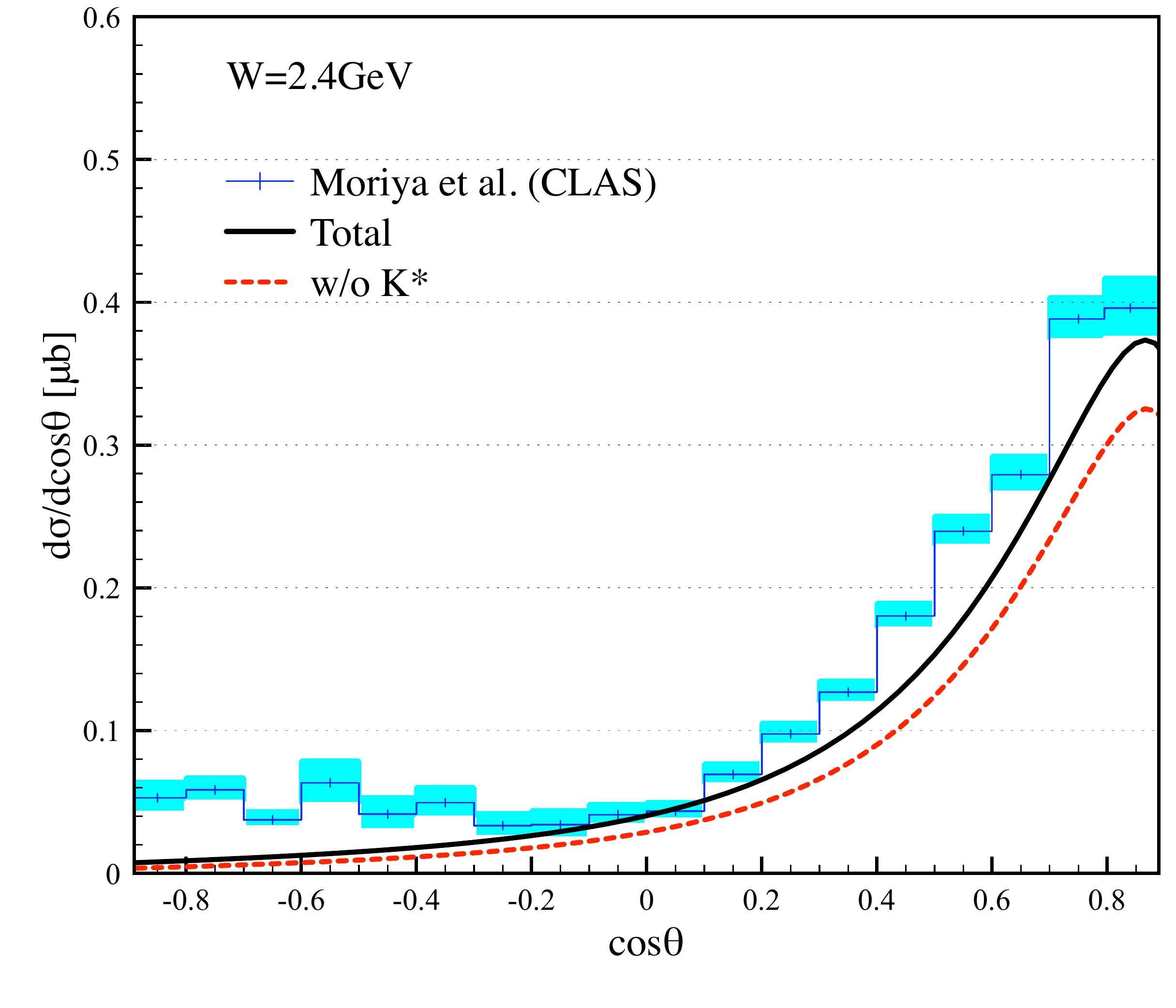}
\includegraphics[width=6cm]{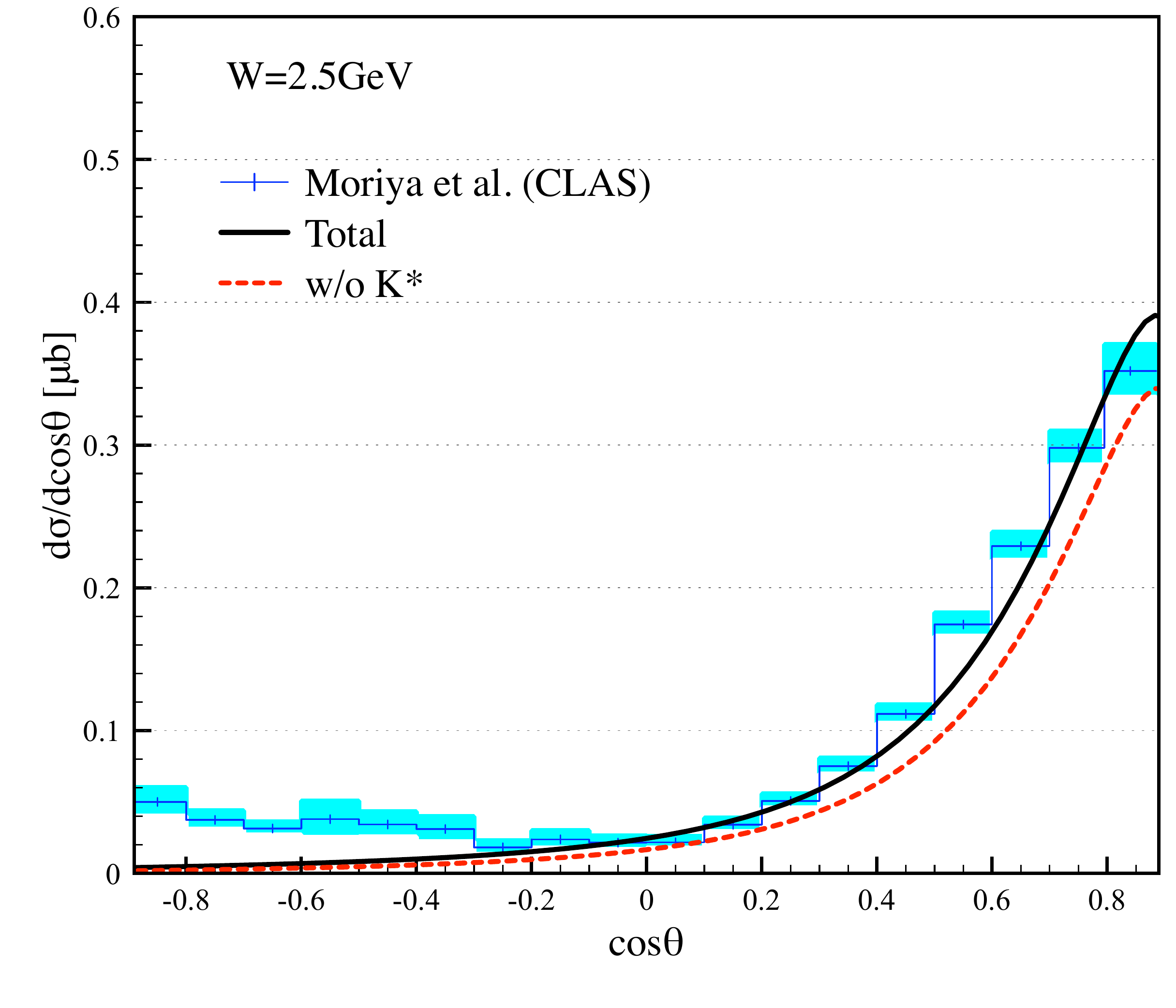}
\end{tabular}
\begin{tabular}{ccc}
\includegraphics[width=6cm]{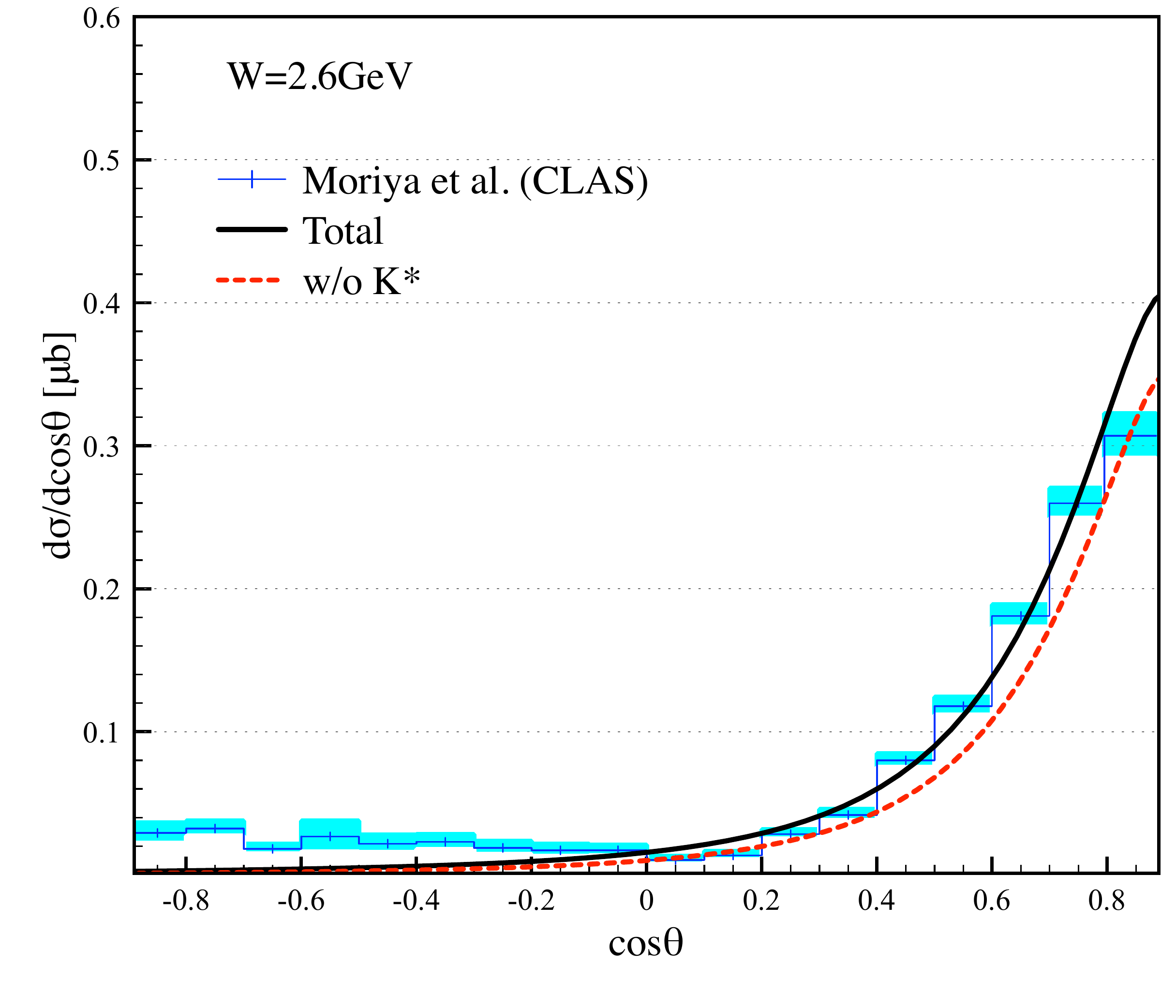}
\includegraphics[width=6cm]{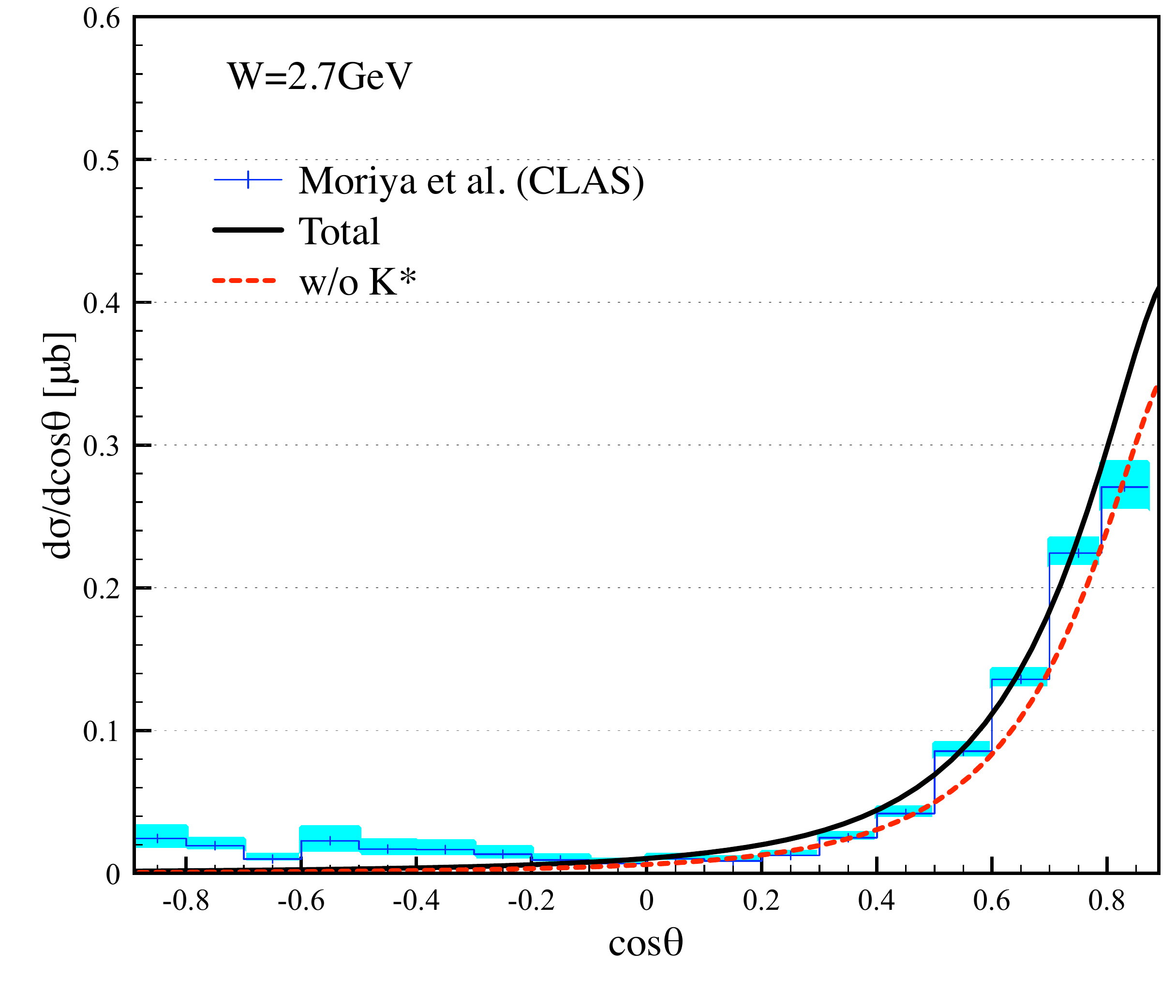}
\includegraphics[width=6cm]{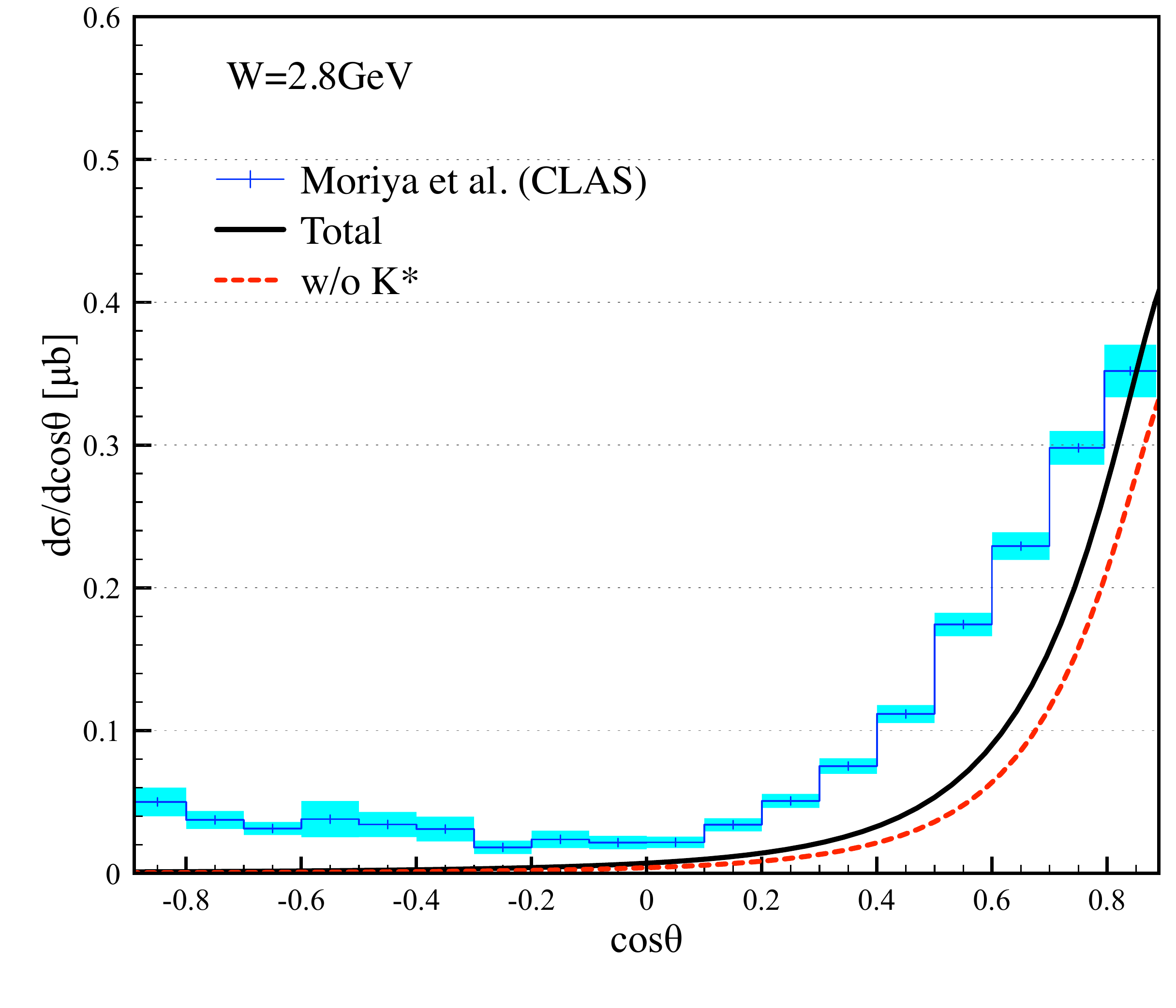}
\end{tabular}
\caption{(Color online) Differential cross section (DCS) $d\sigma/d\cos\theta$ $[\mu b]$ as a function of $\theta$, which indicates the angle for the outgoing kaon in the cm frame for the various cm energies. The experimental data are taken from Ref.~\cite{Moriya:2013hwg}.}       
\label{DCS}
\vspace{1cm}
\begin{tabular}{cc}
\includegraphics[width=7.5cm]{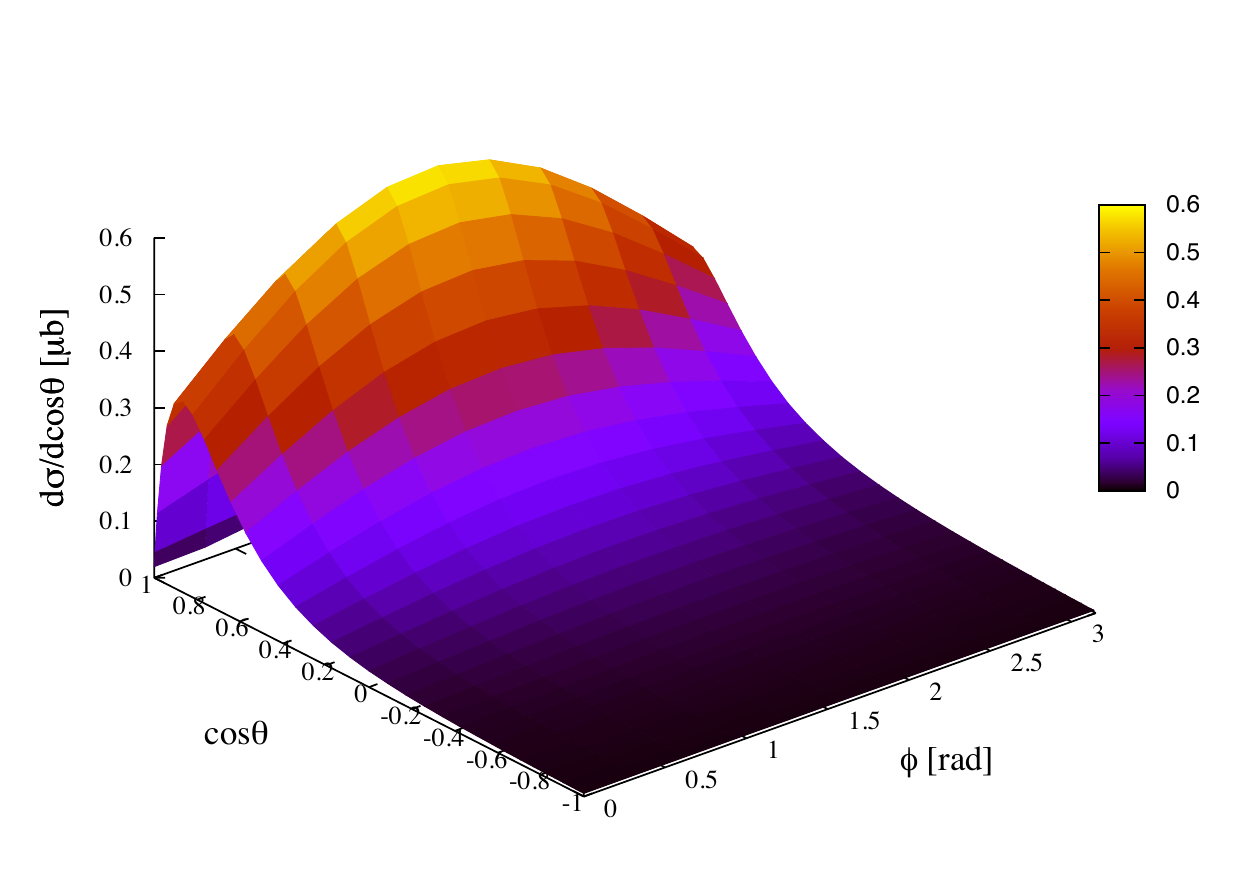}
\includegraphics[width=6cm]{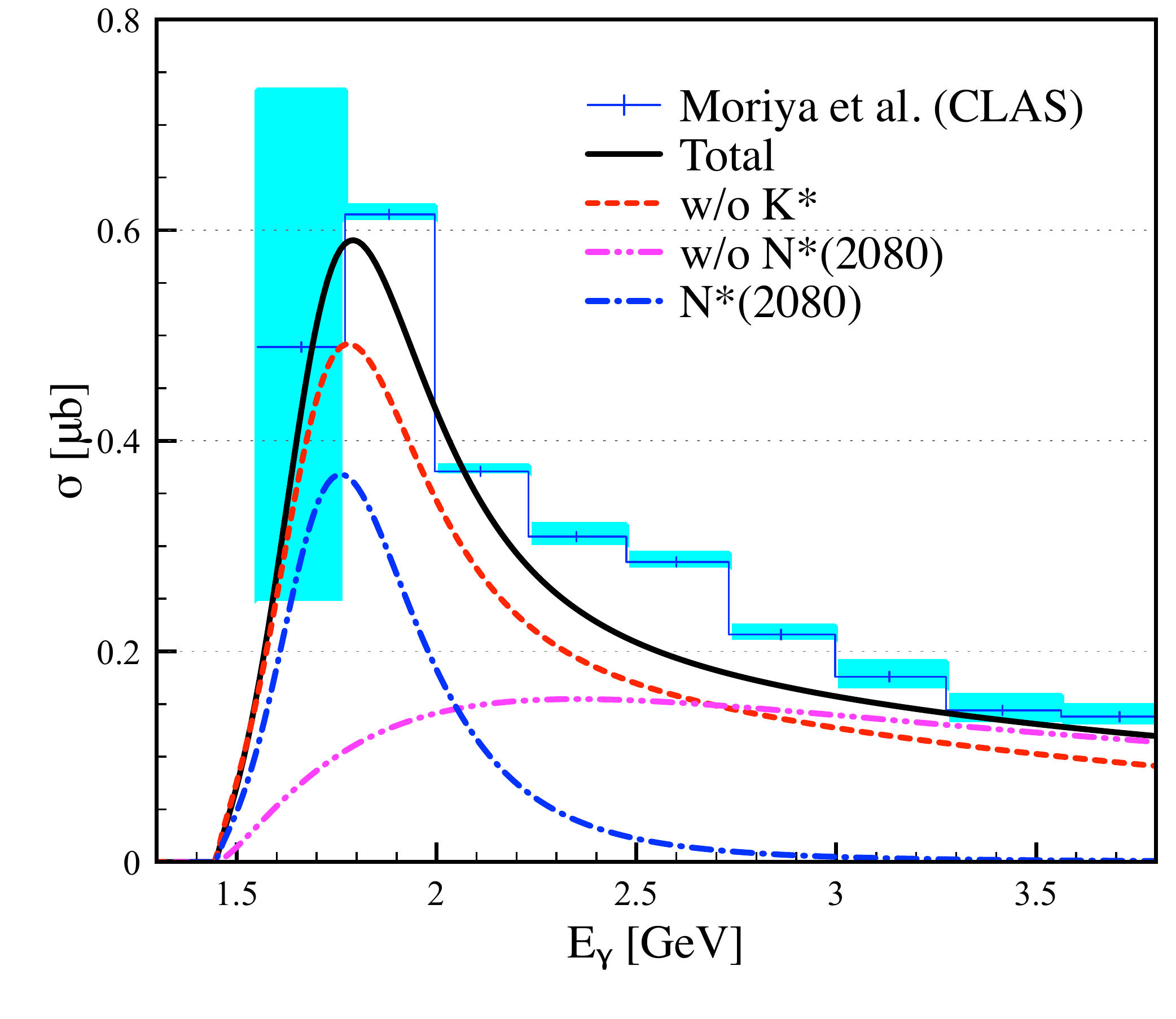}
\end{tabular}
\caption{(Color online) Left: Differential cross section $d\sigma/d\cos\theta$ as a function of the phase angle $\phi$, defined in Eq.~(\ref{eq:BORNAMP}). The cm energy is chosen to be $2.4$ GeV at $M(\pi\Sigma)=1405$ MeV. Right: Total cross  section (TCS) as a function of $E_\gamma$. We consider the Born and resonance ($D_{13}$) contributions at $M(\pi\Sigma)=1405$ MeV. The experimental data are taken from Ref.~\cite{Moriya:2013hwg}.}       
\label{TCS}
\end{figure}

In this Section, we will provide the numerical results and relevant discussions. First, we compute and show the numerical results for the angular dependence, i.e. differential cross section $d\sigma/d\cos\theta$ as a function of $\theta$ for the cm energy range $W=(2.0\sim2.8)$ for $M_{\Lambda^*}=M_{\pi\Sigma}=1405$ MeV. Recently, we have corresponding experimental data from the CLAS collaboration at Jefferson Laboratory (Jlab)~\cite{Moriya:2013hwg}. In Fig~\ref{DCS}, we show the numerical results for it in comparison with the data, in which the shaded area represents the experimental error. Note that the experimental data are the simple sum of the three isospin channels $\pi^{\pm0}\Sigma^{\mp0}$. The numerical results are given by those with the total contributions (solid) and without the $K^*$-exchange contribution (dash). To reproduce the data, we fix the adjustable parameters of the model as listed in Table~\ref{TAB3}.
\begin{table}[b]
\begin{tabular}{>{\centering}m{1.5cm}|
>{\centering}m{1.5cm}|
>{\centering}m{1.5cm}|
>{\centering}m{1.5cm}|
>{\centering}m{1.5cm}|
>{\centering}m{1.5cm}|
>{\centering}m{1.5cm}|
>{\centering}m{1.5cm}|
>{\centering\arraybackslash}m{1.5cm}}
$M_H$&
$M_L$&
$\Gamma_H$
&$\Gamma_L$
&$\phi_H$
&$\phi_L$
&$g_{K^*N\Lambda^*_{H,L}}$
&$M_{N^*{2080}}$
&$\Gamma_{N^*{2080}}$\\
\hline
$1430$ MeV&
$1390$ MeV&
$30$ MeV
&$70$ MeV
&$2.83$
&$0$
&$-2.5$
&$2.0$ GeV
&$230$ MeV\\
\end{tabular}
\label{TAB3}
\caption{Relevant adjustable parameters in the present model.}
\end{table}

The masses of $\Lambda^*_{H,L}$ are almost the same with those estimated from ChUM with a few percent deviations $(M^\mathrm{ChUM}_H,M^\mathrm{ChUM}_L)\approx(1429,1398)$ MeV~\cite{Nam:2003ch}. On the contrary, the full decay width for the higher contribution is quite different from that suggested by ChUM, $\Gamma^\mathrm{ChUM}_H\approx14$ MeV, while the lower one slightly smaller in comparison with $\Gamma^\mathrm{ChUM}_L\approx74$ MeV. The reason for the deviation found in $\Gamma_H$ can be explained by that there are more complicated background effects  in the ChUM calculations, in comparison to our simple setup. The phase angles are chosen to be $\phi_{H,L}=(2.83,0)$, which is almost consistent with the ChUM estimation $\phi_H-\phi_L\approx\pi$~\cite{phaseangle}. The values for $g_{K^*N\Lambda^*_{H,L}}$ is fixed by $-2.5$ for reproducing the data. The mass and full decay width for $D_{13}$ are determined to be $2.0$ GeV and $230$ MeV, respectively. Note that the determined mass is relatively smaller than the expected one $\sim2080$ MeV. With this parameters, we obtain the results in relatively good agreement with the data. We observe that, with the $K^*$-exchange contribution, the strength of the cross section increases obviously. All the curves show strong forward-scattering enhancements, due to the $K$-exchange in the $t$ channel dominantly. The experimental data shows sizable enhancement in the cross section in the backward-scattering region. Although this enhancement can be explained by possible $u$-channel resonant contributions, as discussed previously, $\Lambda(1670)$ does not meet this requirement. The inclusion of $\Sigma(1385)$ could help this, but we would like to leave it as a future work. In the left panel of Fig.~\ref{TCS}, we show the numerical results for the differential cross section as a function of $\cos\theta$ and $\phi\equiv(\phi_H-\phi_L)$. It turns out that the dependence on $\phi$ is relatively smooth. Once all the parameters fixed, we compute the total cross section  as a function of the photon energy $E_\gamma$ and show the numerical results in the right panel of Fig.~\ref{TCS}. The experimental data are taken again from Ref.~\cite{Moriya:2013hwg}. We show the theoretical curves with the total contributions (solid), without $K^*$ (dash), without $D_{13}$ (dot-dash), and $D_{13}$ only (dot-dot-dash). We observe that the $D_{13}$ contribution dominates in the vicinity near the threshold, whereas it diminishes stiffly beyond $E_\gamma\approx2.0$ GeV then the usual Born-term contributions remain. The $K^*$-exchange contribution again provides strength enhancement over the photon-energy region that we focus on. The experimental data are reproduced qualitatively well. 

\begin{figure}[t]
\begin{tabular}{ccc}
\includegraphics[width=6cm]{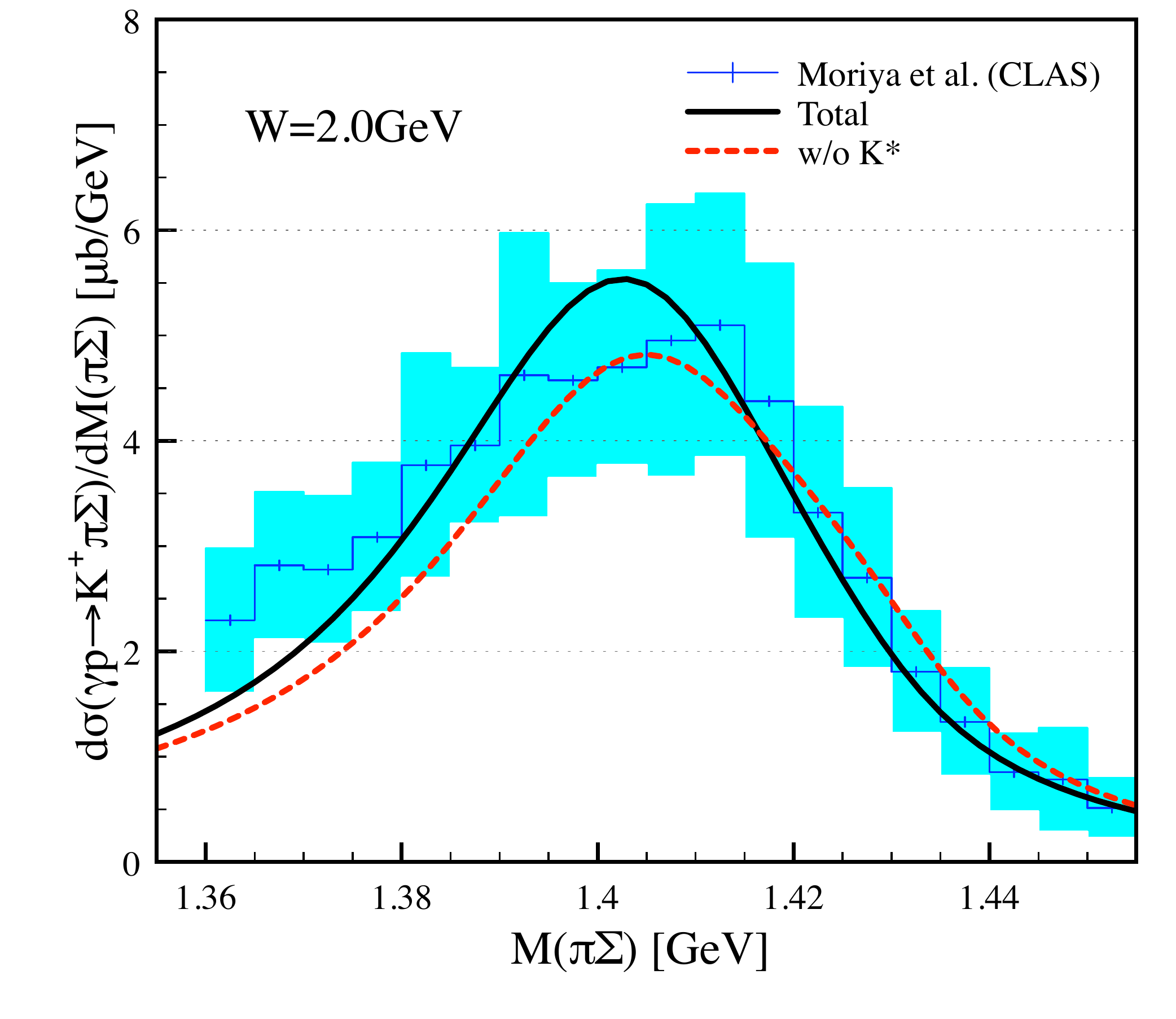}
\includegraphics[width=6cm]{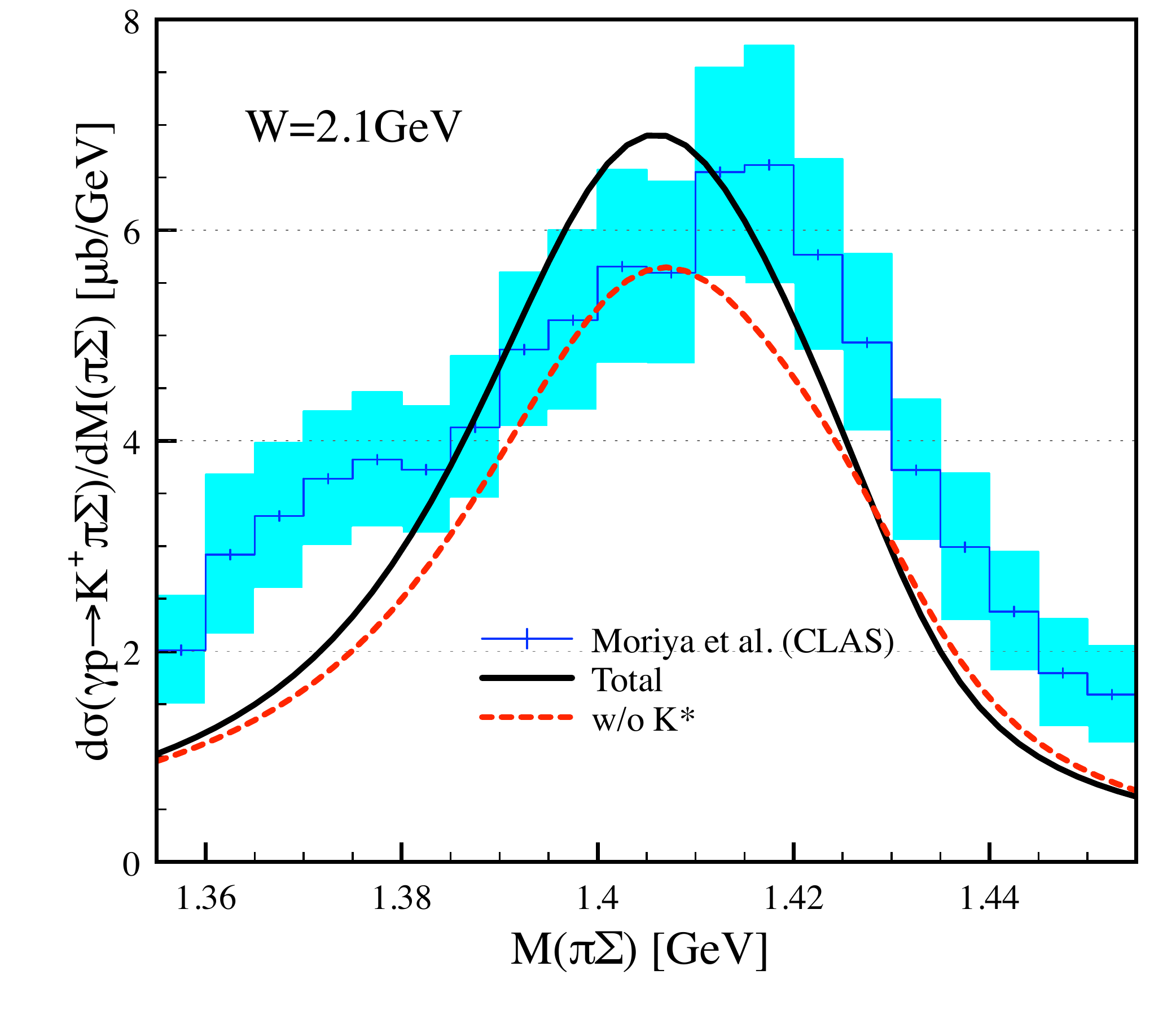}
\includegraphics[width=6cm]{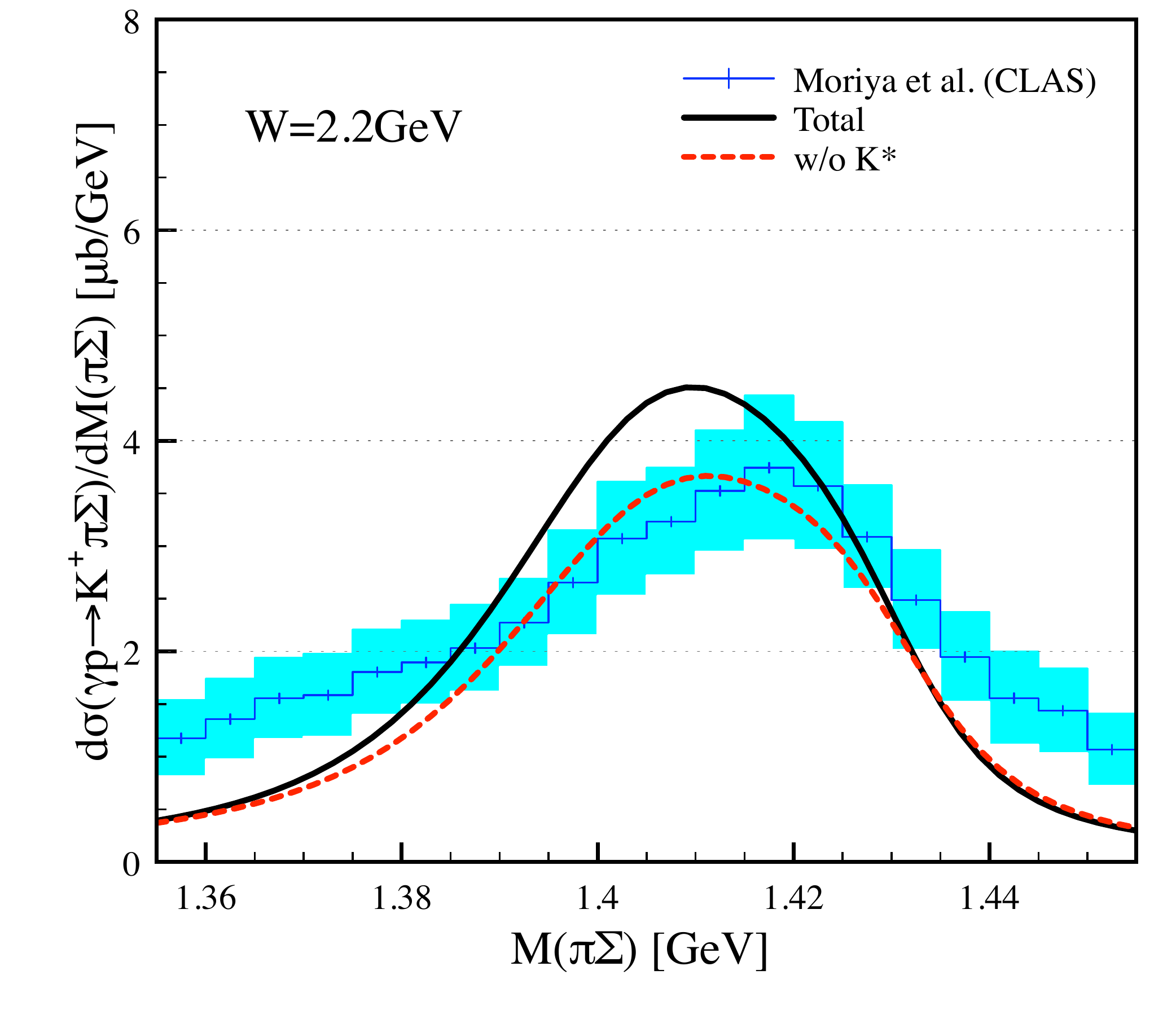}
\end{tabular}
\begin{tabular}{ccc}
\includegraphics[width=6cm]{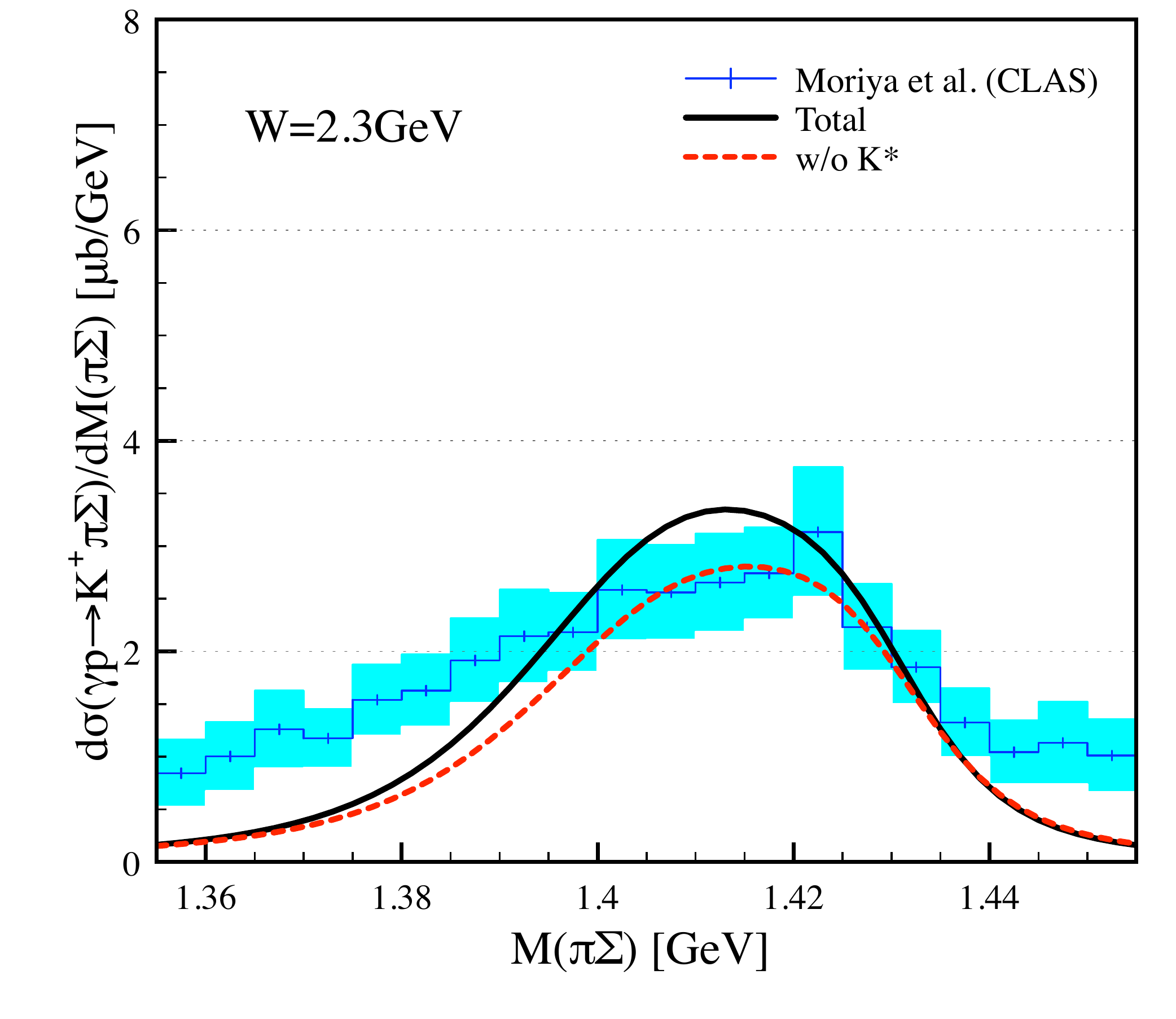}
\includegraphics[width=6cm]{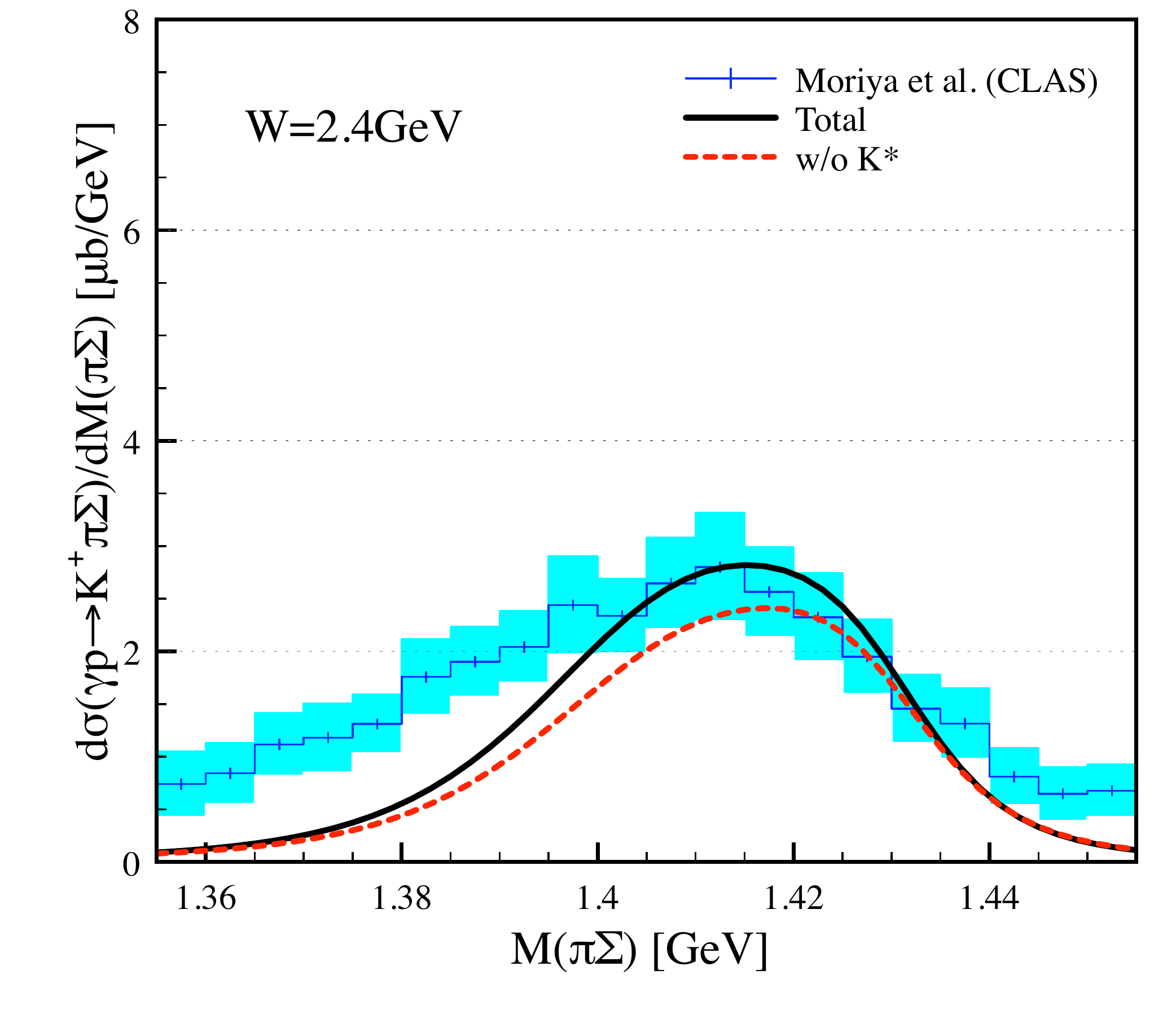}
\includegraphics[width=6cm]{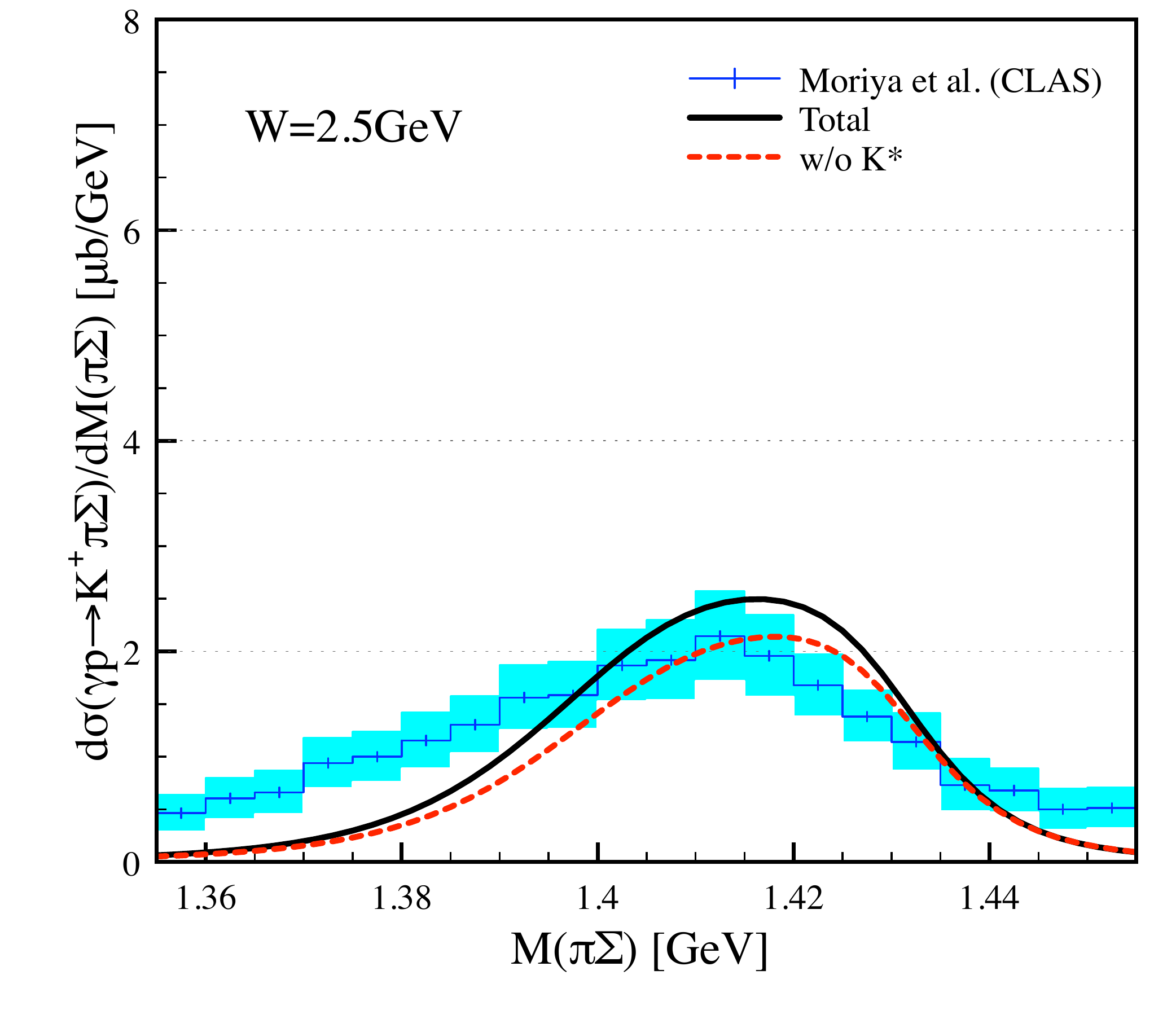}
\end{tabular}
\begin{tabular}{ccc}
\includegraphics[width=6cm]{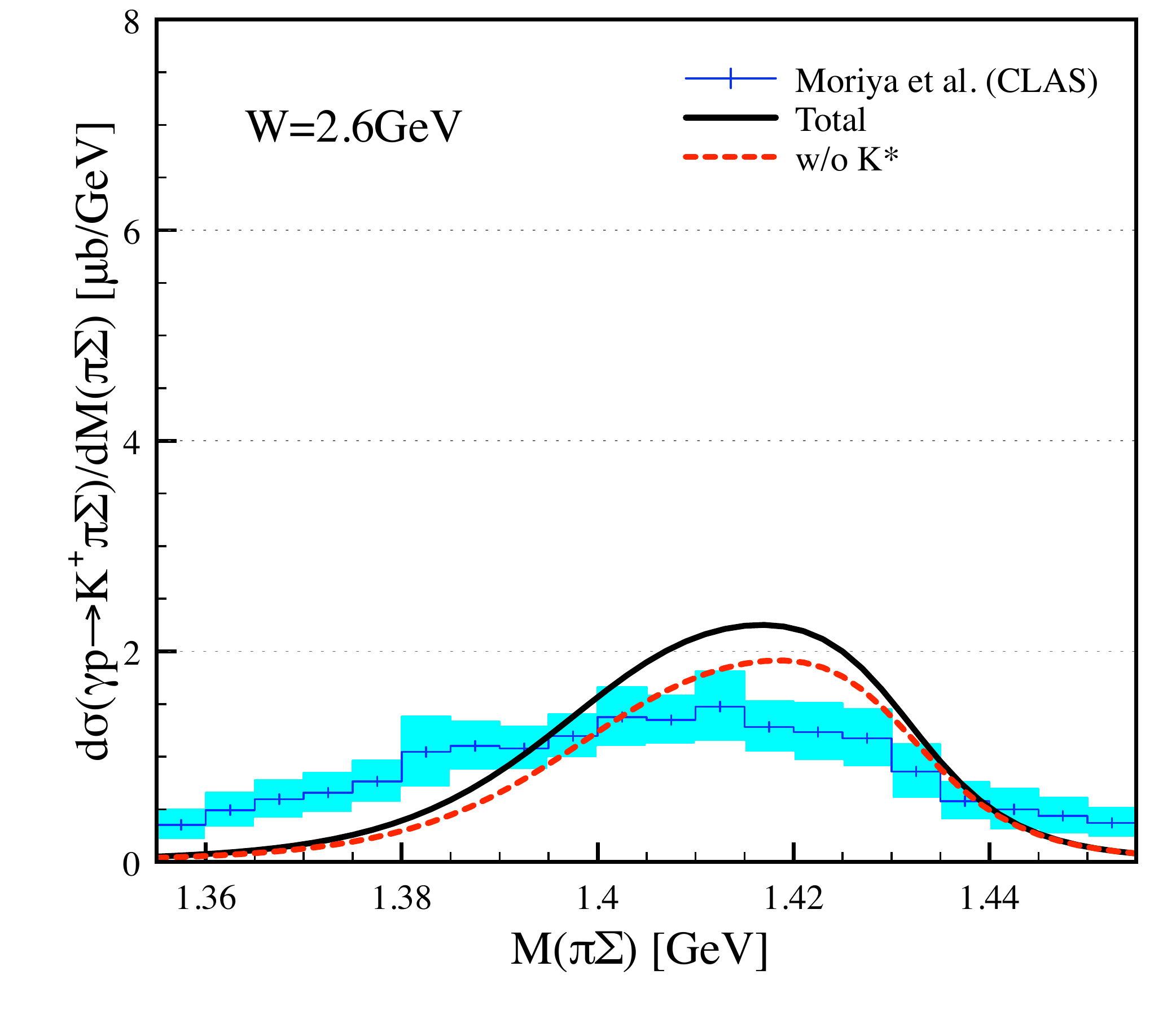}
\includegraphics[width=6cm]{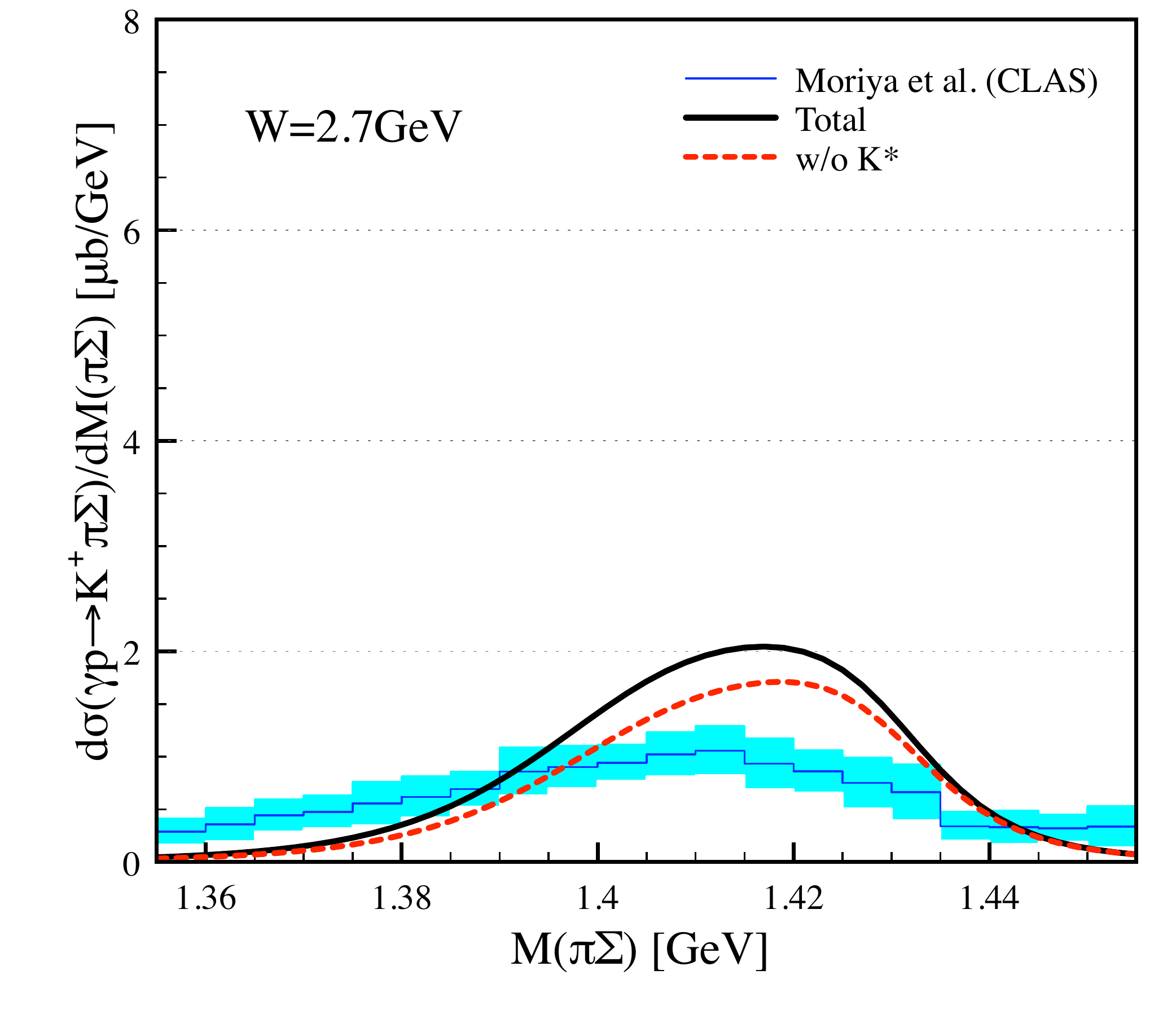}
\includegraphics[width=6cm]{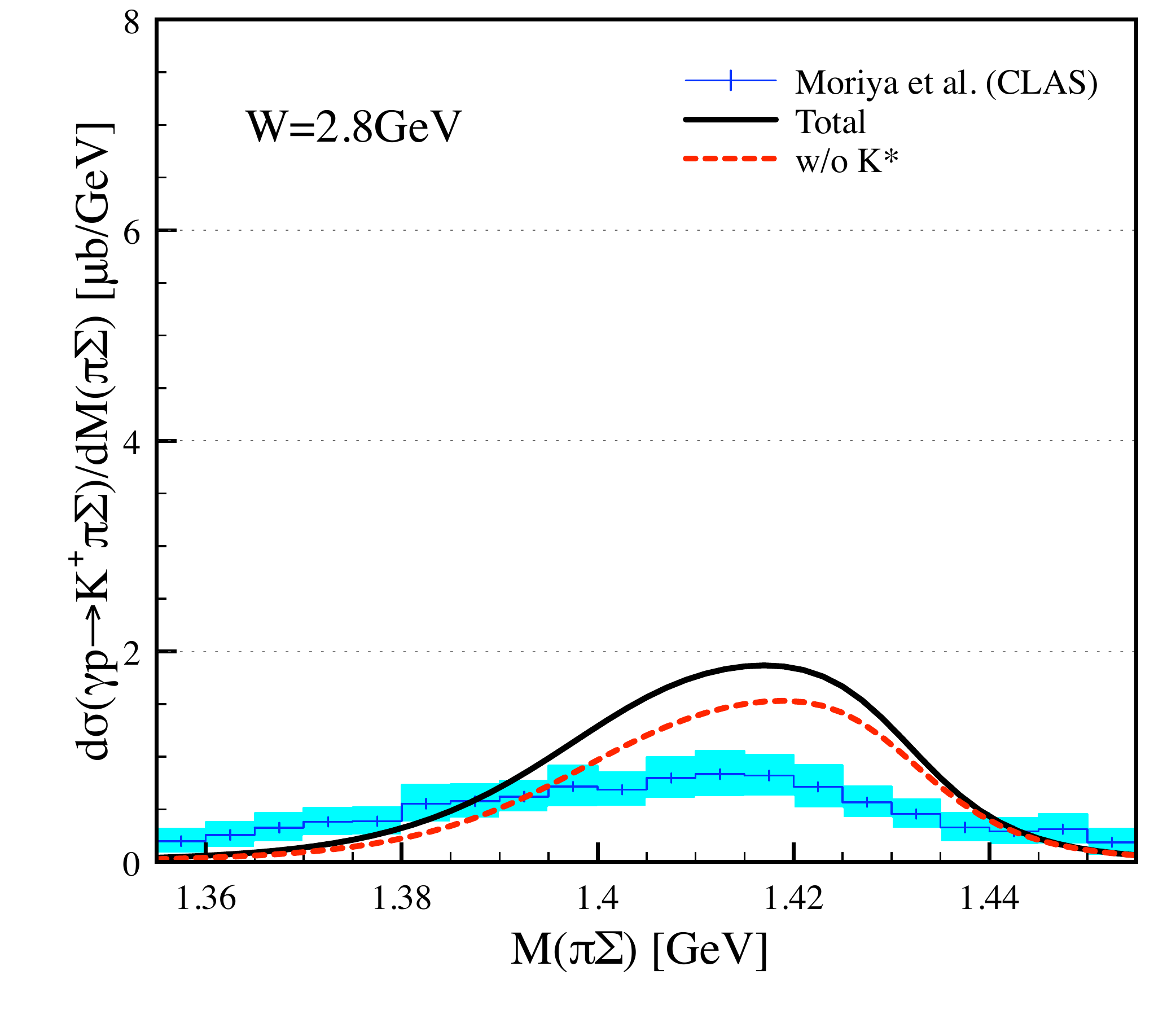}
\end{tabular}
\caption{(Color online) Left: Differential cross section $d\sigma(\gamma p\to K^+\pi\Sigma)/dM(\pi\Sigma)$ as a function of $M(\pi\Sigma)$ for $W=(2.0\sim2.8)$ GeV with (solid) and without (dash) the $K^*$-exchange contribution.}       
\label{INVMASS}
\vspace{1cm}
\begin{tabular}{cc}
\includegraphics[width=6cm]{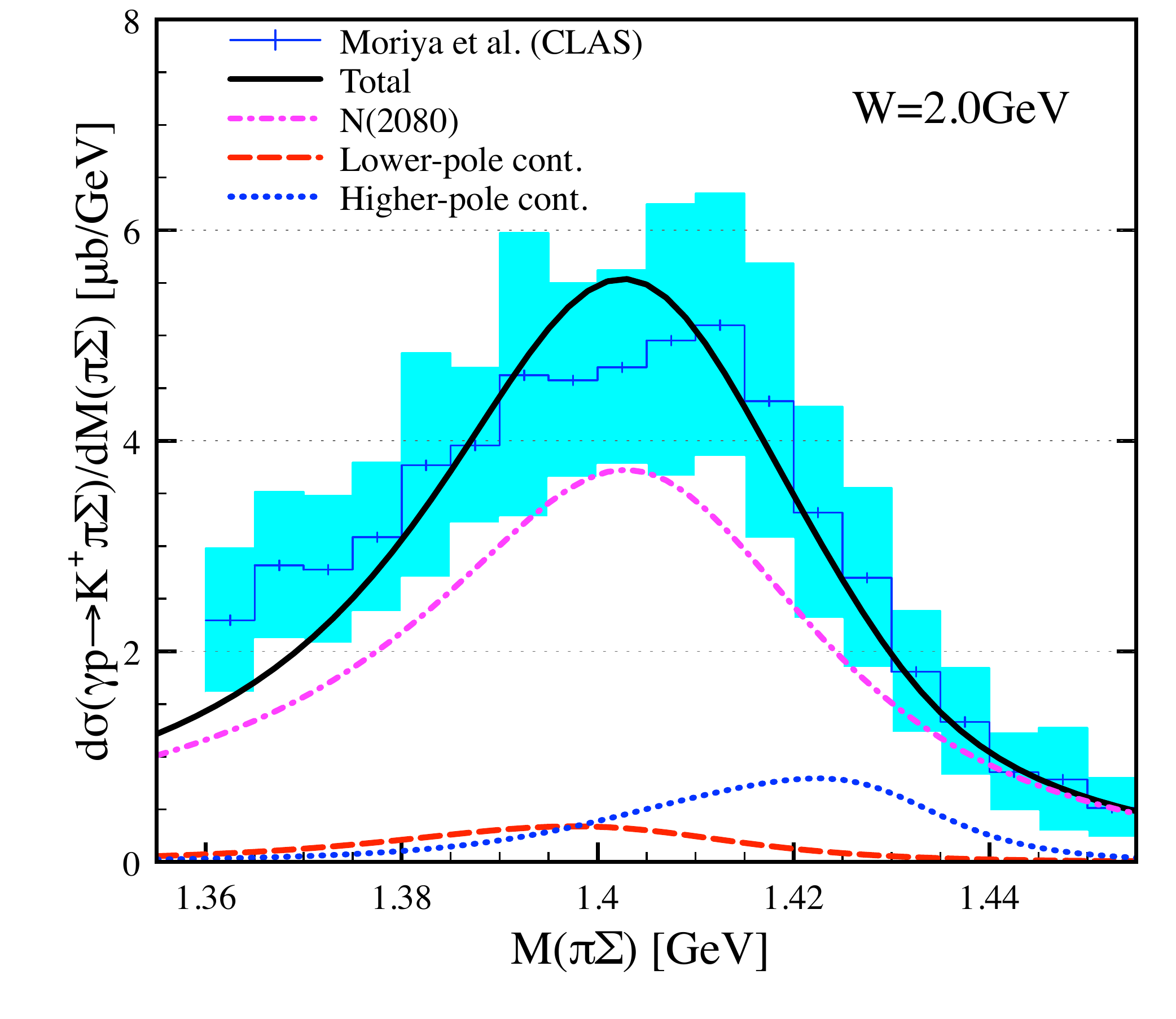}
\hspace{1cm}
\includegraphics[width=6cm]{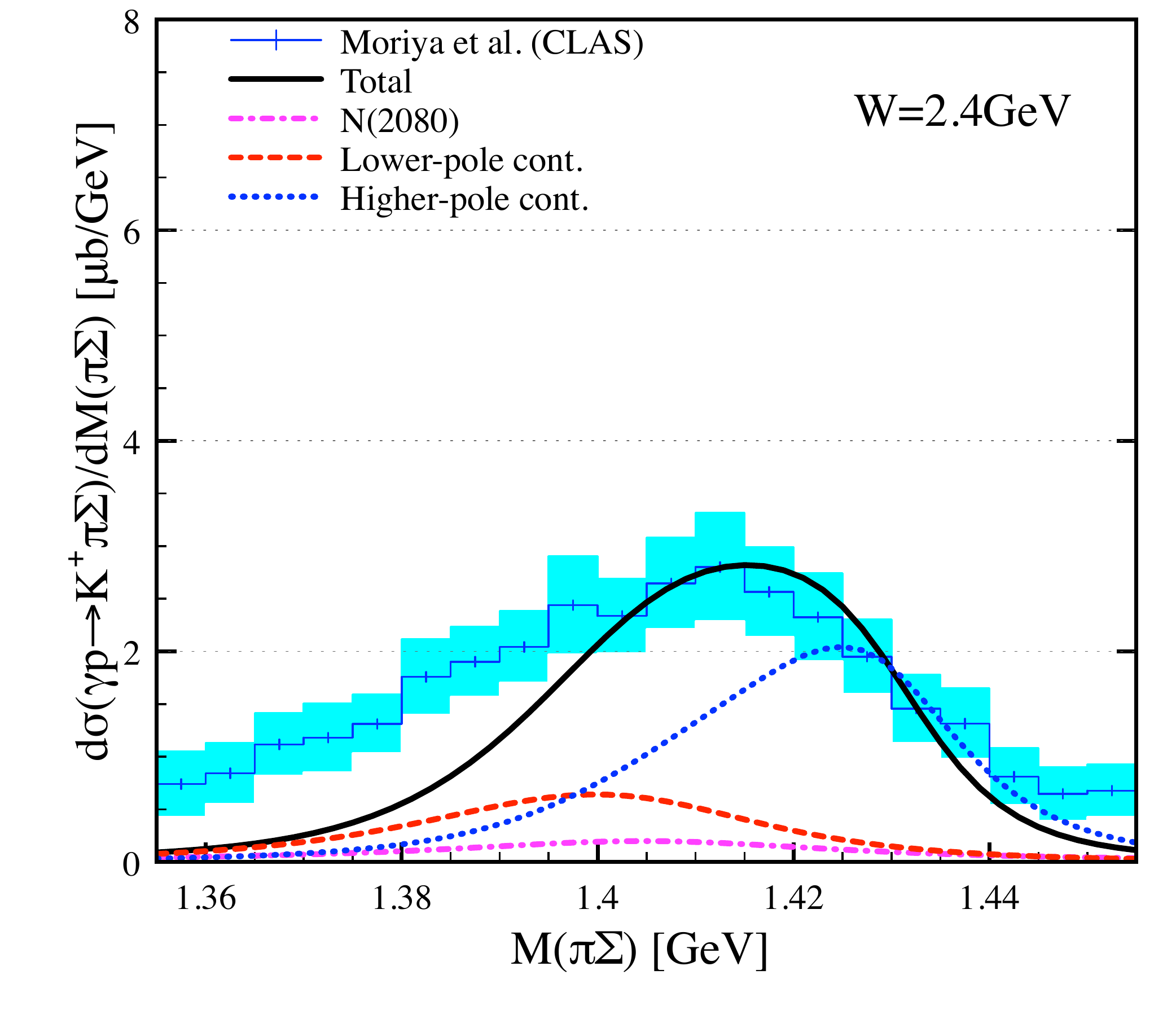}
\end{tabular}
\caption{(Color online) Left: Differential cross section $d\sigma(\gamma p\to K^+\pi\Sigma)/dM(\pi\Sigma)$ as a function of $M(\pi\Sigma)$ at $W=2.0$ GeV, indicating total (solid), $N(2080)$ (dot-dash), lower-pole (long-dash), and higher-pole (dot) contributions, separately. The experimental data are taken from Ref.~\cite{Moriya:2013eb}. Right: The same plot for $W=2.4$ GeV.}       
\label{INVMASSANAL}
\end{figure}

One of the distinguished features of the present model must be that the $\Lambda(1405)$ mass is treated as a dynamical variable, i.e. the $\pi$-$\Sigma$ invariant mass ($M_{\pi\Sigma}$). Hence, it is interesting to extract the $\pi$-$\Sigma$ invariant-mass distribution from the two-body reaction-process calculations. If we assume that the Dalitz process $\gamma p\to K^+\pi\Sigma$ is saturated by the $\Lambda^*_{H,L}$ intermediate states, and their decay widths are small enough comparing to their masses, the invariant-mass distribution for the Dalitz process can be defined with the unpolarized two-body process $\sigma_{\gamma p\to K^+\Lambda^*}$ as a function of $M_{\pi\Sigma}$:
\begin{eqnarray}
\label{eq:3to2}
\frac{d\sigma_{\gamma p\to K^+\pi\Sigma}}{dM_{\pi\Sigma}}\approx\frac{2M_{\Lambda^*}\,M_{\pi\Sigma}}{\pi}\frac{\sigma_{\gamma p\to K^+\Lambda^*}\,\Gamma_{\Lambda^*\to\pi\Sigma}}{(M^2_{\pi\Sigma}
-M^2_{\Lambda^*})^2+M^2_{\Lambda^*}\Gamma^2_{\Lambda^*}}.
\end{eqnarray}
Here, $\Lambda^*\equiv\Lambda(1405)$ and $M_{\pi\Sigma}\equiv M(\pi\Sigma)$. The explicit derivation of Eq.~(\ref{eq:3to2}) is given Appendix in detail. Note that we also assume for Eq.~(\ref{eq:3to2}) that the interference between the $\Lambda^*_{H,L}$ intermediate process and the $K^*$-meson pole contribution, in which $K^*$ decays into $K^+$ and $\pi$, is negligible.  Using Eq.~(\ref{eq:3to2}) and the total cross section results for $\sigma_{\gamma p\to K^+\Lambda^*}$, the computed invariant-mass distribution is given in Fig.~\ref{INVMASS} for $W=(2.0\sim2.8)$ GeV and $M_{\pi\Sigma}=(1.355\sim1.455)$ MeV. The experimental data are taken from Ref.~\cite{Moriya:2013eb}. Again, we show the experimental data as a simple sum of the three $\pi$-$\Sigma$ isospin channels. The solid and dash lines indicate the numerical results with and without the $K^*$-exchange contribution, respectively. As $W$ increases, the peak of the curves tend to be slightly shifted to the higher $M_{\pi\Sigma}$ value. In other words, for the lower-energies, the distribution is more symmetric inside the mass window $M_{\pi\Sigma}=(1.355\sim1.455)$ MeV, and the theory explains this tendency qualitatively well. Most obvious deviations in comparison to the experimental data are the underestimations below $M_{\pi\Sigma}\approx1.4$ GeV, and the overestimations beyond $W=2.6$ GeV. We can understand the first by the absence of the possible $\Sigma(1385)$ contribution in the present work. As for the second, there can be more complicated destructive interference between the presently considered contributions and those ignored here.  One of the possible cures for this overestimations must be the inclusion of the Regge contributions in the $t$ channel for the relatively higher-energy region. We would like to leave this for the future work. 

\begin{figure}[t]
\begin{tabular}{cc}
\includegraphics[width=8.5cm]{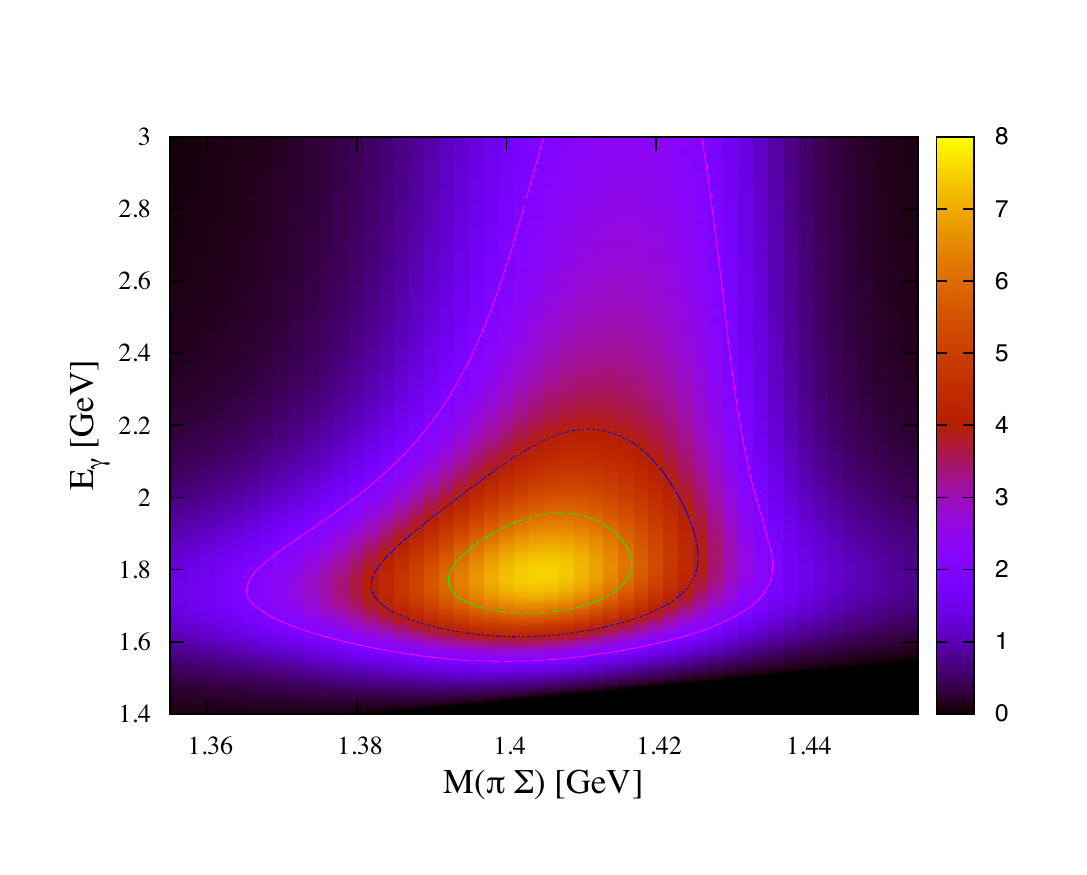}
\includegraphics[width=8.5cm]{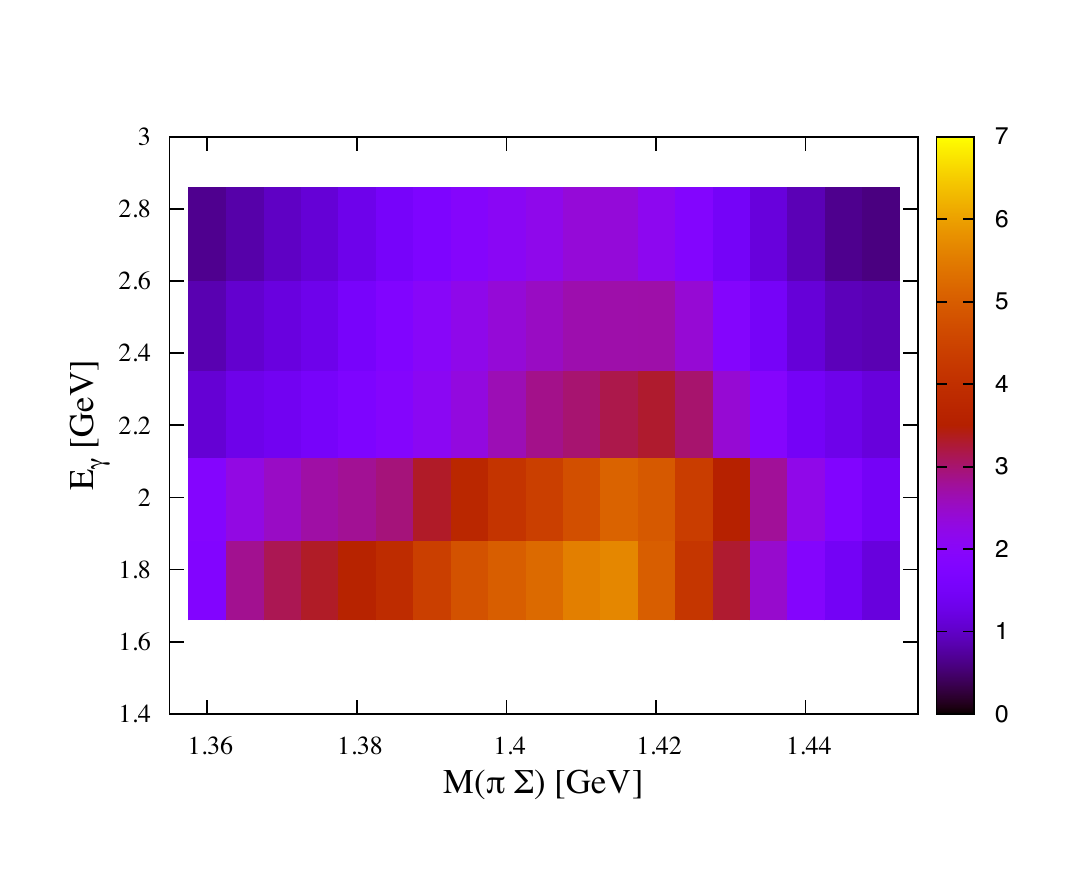}
\end{tabular}
\caption{(Color online) Left: Theoretical results for the differential cross section $d\sigma(\gamma p\to K^+\pi\Sigma)/dM(\pi\Sigma)$ as a function of $E_\gamma$ and $M(\pi\Sigma)$. Right: The same plot with the experimental data taken from Ref.~\cite{Moriya:2013eb}, ignoring the errors. One can observe the tilted-triangle-shape (${\it \Delta}$-shape) distribution by seeing the outmost contour line in the left panel. A similar tendency can also be found in the experimental data. }       
\label{INVMASS3D}
\vspace{1cm}
\begin{tabular}{cc}
\includegraphics[width=7cm]{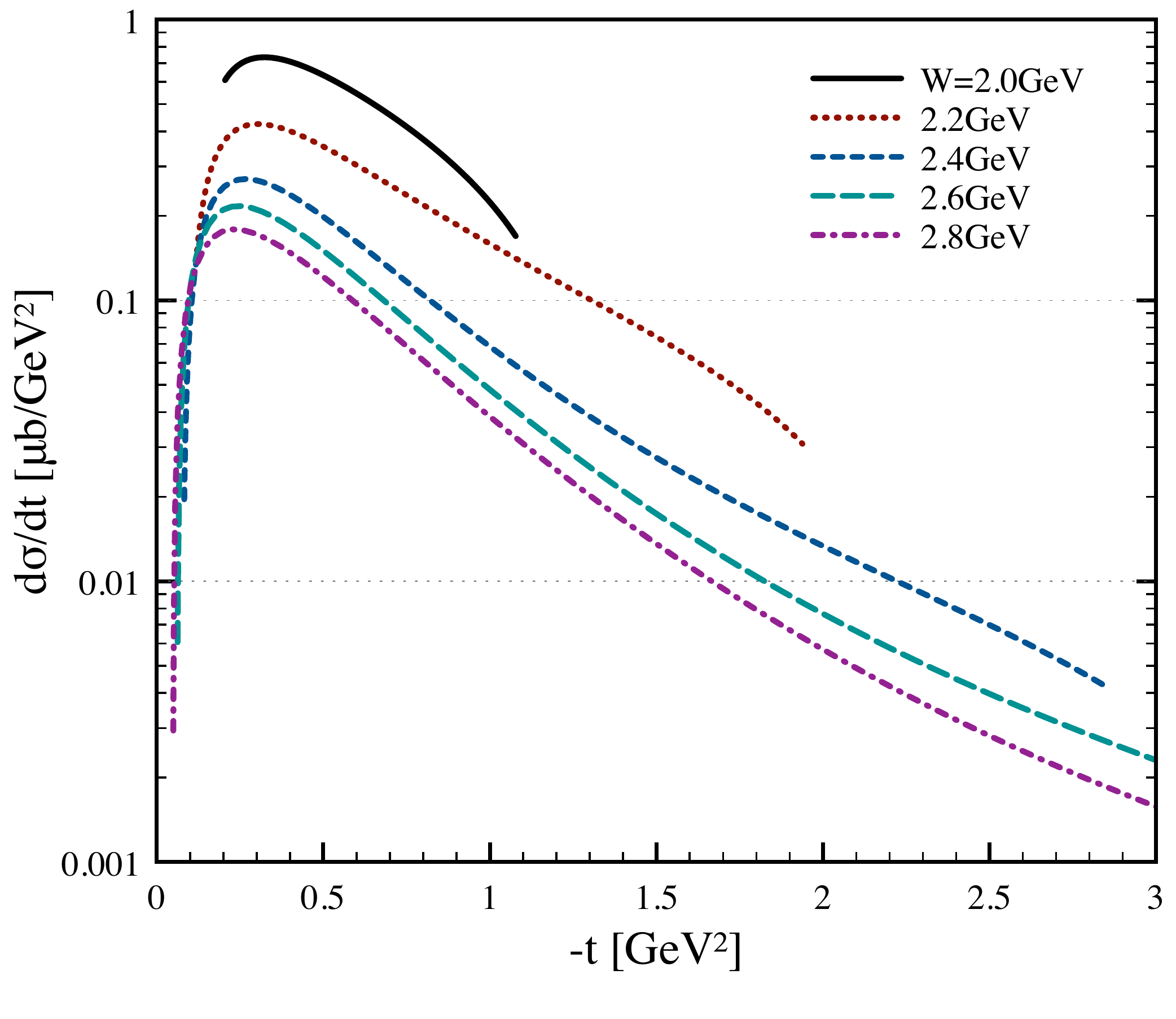}
\hspace{1cm}
\includegraphics[width=7cm]{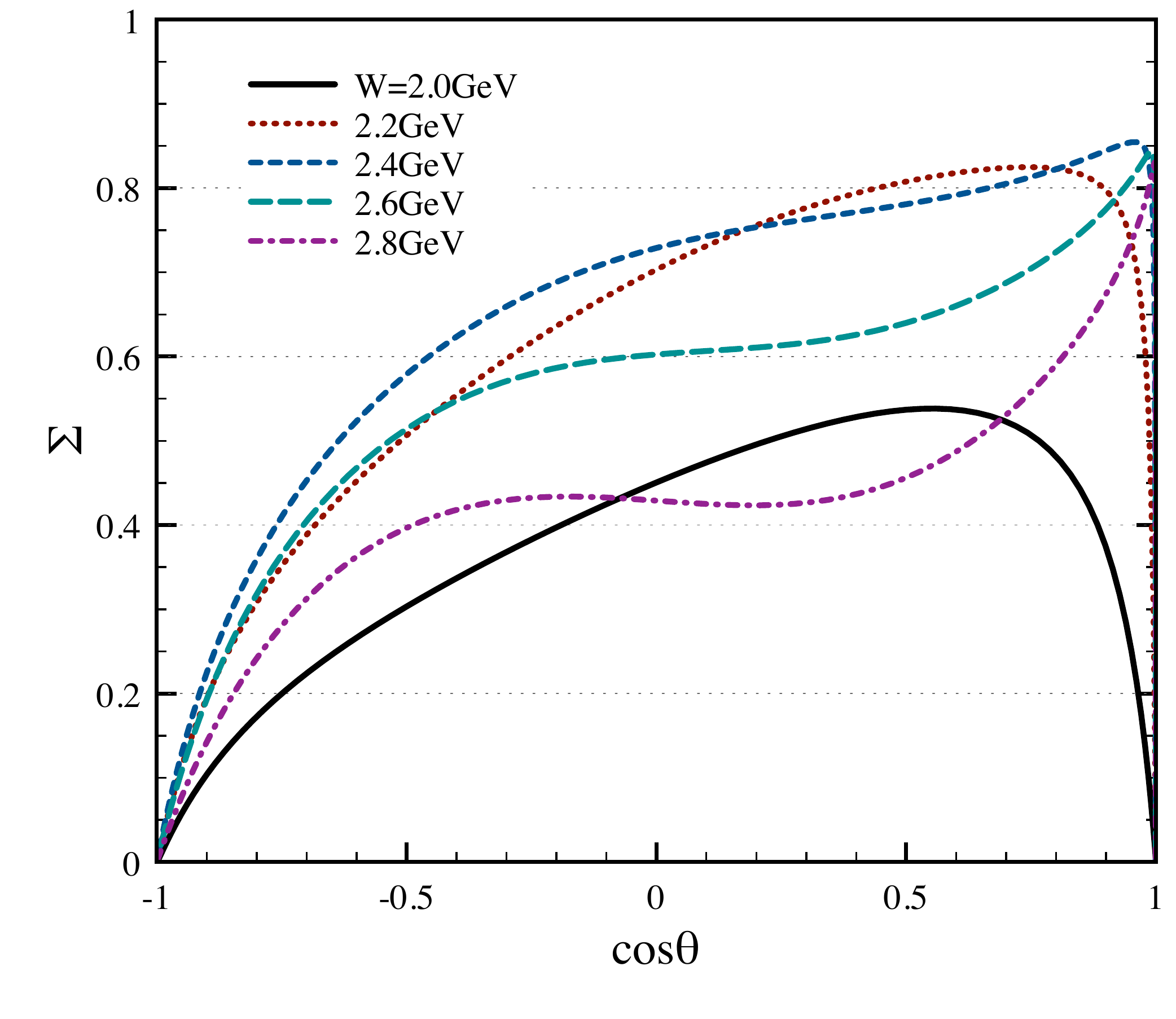}
\end{tabular}
\caption{(Color online) Left: $t$-channel momentum transfer $d\sigma/dt$ as a function of $-t$ for various cm energies. Right: Photon-beam asymmetry $\Sigma$ as a function of $\cos\theta$ for various c.m. energies.}       
\label{OTHER}
\end{figure}

In order to understand the peak shift or becoming asymmetric in the invariant-mass distributions in detail, we decompose them into the relevant contributions and show the results in Fig.~\ref{INVMASSANAL}. To see the tendency clearly, we choose two typical energies, $W=(2.0,2.4)$ GeV in the (left, right) panels. As for $W=2.0$ GeV, the $D_{13}$ contribution (dot-dash) becomes dominant, where as the $\Lambda^*_{H,L}$ ones are considerably small. Hence, due to Eq.~(\ref{eq:3to2}), the distribution has a peak near $M_{\pi\Sigma}\approx1405$ MeV and looks symmetric within the window. As the energy grows, the resonance contribution gets diminished, and the $\Lambda^*_{H,L}$ ones start to prevail. Hence, the two-pole structure gives asymmetric distribution as shown in the right panel. This observation can be seen more clearly in Fig.~\ref{INVMASS3D}. There, we plot the invariant-mass distribution as a function of $E_\gamma$ and $M_{\pi\Sigma}$. The theory and experiment are given in the left and right panels, respectively. For the experimental data, we ignore the error here. By seeing the outmost contour in the left panel, we find a tilted-triangle shape ($\it \Delta$-shape) distribution. A similar tendency can be seen even in the experimental data in the right panel. Once again, this observation can be understood by that the dominant resonance contribution near the threshold, and the generic Born ones becomes of importance as the energy increases. Moreover, the two-pole structure with the different strengths gives the asymmetric distribution. 

Finally, we want to discuss and provide theoretical results for two more physical observables, i.e. $t$-channel momentum transfer $d\sigma/dt$ and the photon-beam asymmetry $\Sigma$. In the left panel of Fig.~\ref{OTHER}, we show the numerical results for $d\sigma/dt$ as a function of $-t$ for $W=(2.0\sim2.8)$ GeV . It turns out that the curves behave similarly with other pseudo-scalar meson photoproductions in general.  A peculiar feature is that, due to the $D_{13}$ contribution near the threshold, the curves below $W=2.2$ GeV are considerably larger than those beyond it. The photon-beam asymmetry is defined in the present work by
\begin{equation}
\label{eq:BA}
\Sigma=\frac{\frac{d\sigma}{d\Omega}_\perp-\frac{d\sigma}{d\Omega}_\parallel}{\frac{d\sigma}{d\Omega}_\perp+\frac{d\sigma}{d\Omega}_\parallel},
\end{equation}
where the subscripts $\perp$ and $\parallel$ denote that the photon polarization is perpendicular and parallel to the reaction plane, on which four momenta of all the particles reside. In our definition of $\Sigma$, the $K$-exchange contribution in the $t$ channel gives $\Sigma\approx1$, whereas the $K^*$-exchange one $\Sigma<0$ as a function of $\cos\theta$. For the lower-energy region around the $D_{13}$ mass, i.e. $W\lesssim2.1$ GeV, there are destructive interferences between the $K$-exchange  and $D_{13}$ contributions, giving $\Sigma\approx0.5$. As the energy increases to $W\approx2.4$ GeV, the $D_{13}$ contributions gets diminished, then the shape of $\Sigma$ approaches to that of the $K$-exchange one $\Sigma\to1$. When the energy goes beyond $W\approx2.4$ GeV, the $K^*$-exchange in the $t$ channel becomes effective, and provides destructive interference with the $K$-exchange one, resulting in $\Sigma\approx0.4$ averagely for $W=2.8$ GeV with the negligible $D_{13}$ contribution.

\section{Summary, conclusion, and perspectives}
We have studied the $\Lambda(1405)\equiv\Lambda^*$ photoproduction, considering its two-pole structure as suggested by the ChUM calculations. For this purpose, we developed a simple two-body  process model, based on the effective Lagrangian approach. In this model, the hypothetical states of $\Lambda^*_{H,L}$, suggested by ChUM, are coupled to the physically measured $\pi$-$\Sigma$  state, denoted by $\Lambda^*$, whose mass is assigned as the $\pi$-$\Sigma$ invariant mass $M_{\pi\Sigma}$. By doing this, the model mimics the Dalitz process, i.e. $\gamma p \to K^+\pi\Sigma$, approximately with the two-body reaction process $\gamma p \to K^+\Lambda^*$. Relevant model parameters were determined by various theoretical and experimental information, such as the results from ChUM. Especially, the coupling strengths between $\Lambda^*_{H,L}$ and $\Lambda^*$ were determined by the meson-baryon loop with help of the on-shell factorization and dimensional regularization. We considered the various contributions, including the $D_{13}$ one, with the gauge-invariant form factor scheme. We computed various physical quantities and list important observations as follows:
\begin{itemize}
\item All the adjustable model parameters are determined to reproduce the recent CLAS experiment data, and it turns out that they are not the same but relatively similar to those estimated from ChUM. The deviations from the ChUM estimations can be understood by different backgrounds considerations between the present model and ChUM coupled-channel calculations. We also note that the $K^*$-exchange contribution is finite.
\item The angular ($d\sigma_{\gamma p\to K^+\Lambda^*}/d\cos\theta$) and energy ($\sigma_{\gamma p\to K^+\Lambda^*}$) dependences of the cross sections at $M_{\Lambda^*}=M_{\pi\Sigma}=1405$ MeV are reproduced qualitatively well in comparison to the experimental data. We find that the nucleon resonance contribution from $D_{13}$ dominates the low-energy region near the threshold, and the $K$-exchange in the $t$ channel plays important role for the wide energy range. The $K^*$-exchange contribution gives overall enhancement to the cross sections in general. 
\item The invariant-mass distribution ($d\sigma_{\gamma p\to K^+\pi\Sigma}/dM_{\pi\Sigma}$) is computed as a function of $M_{\pi\Sigma}$, showing qualitative agreement with the experimental data. Focusing on the window $M_{\pi\Sigma}=(1.355\sim1.455)$ MeV, the distribution is relatively symmetric, due to the dominant $D_{13}$ contribution near the threshold. As the energy grows, together with the diminishing $D_{13}$ contribution, the distribution becomes shifted by the asymmetry generated by the two-pole structure of $\Lambda(1405)$. Again, the $K^*$-exchange contribution shows overall enhancement to the invariant-mass distribution.
\item The $t$-channel momentum transfer ($d\sigma_{\gamma p\to K^+\Lambda^*}/dt$) and photon-beam asymmetry $(\Sigma)$ are also studied theoretically. In the beam asymmetry, it clearly shows the destructive interference between the $D_{13}$ and $K$-exchange contributions near the threshold. As the energy increases, the $K^*$-exchange contribution comes into play significantly. For the energy range $W=(2.0\sim2.8)$ GeV, the theory shows $\Sigma\sim0.5$ averagely. 
\item Although the theoretical results shows qualitatively good agreement with the experimental data, we note that some issues are not described well in the present work: 1) The experimental data for the angular dependence indicates sizable $u$-channel contributions, we can not reproduce them even with the inclusion of $\Lambda(1670)$. 2) The inclusion of $\Sigma(1385)$ contribution, which can couple to $\Lambda^*$ can give better descriptions for the invariant-mass distribution for $M_{\pi\Sigma}\lesssim1.4$ GeV. 3) The invariant-mass distribution is overestimated for $W\gtrsim2.6$ GeV. This can indicate a necessity for the high-energy modification of the present model, such as the Regge-trajectory for the $t$-channel contributions.   
\end{itemize}

In consequence, the present model inspired by the two-pole structure of $\Lambda(1405)$ reproduces the experimental data qualitatively well  with helps of ChUM. Note that describing invariant-mass distribution with the two-body reaction process is one of the specific features of the present model. Hence, the present model is quite simple and useful to analyze various Dalitz processes in an easier way. Thus, the successful description of the experimental data within the present model calculations with the ChUM information can be considered supporting the two-pole structure of $\Lambda(1405)$. Nonetheless, it is still difficult to make a conclusive statement for the two-pole structure of $\Lambda(1405)$ within the present model study, since the single-pole scenario may lead us to the similar consequence by reproducing the data. We want to extend the present model calculations by including the $\Sigma(1385)$ intermediate state, Regge trajectories in the $t$ channel, and so on. Related works are under progress and will appear elsewhere. 

\section*{Acknowledgment}
The authors thank D.~Jido, T.~Hyodo, K.~Moriya, R.~Schumacher, K.~H.~Woo, and K.~T.~Nam for the fruitful discussions and helps. SiN is grateful to the support by JSPS KAKENHI Grant Number 25400254 (D. Jido), during his stay at Tohoku University, Japan. This work was supported in part by the Korea Foundation for the Advancement of Science and Creativity (KOFAC) grant funded by the Korea government (MEST) (20142169990) and by the NRF grant funded by MEST (Center for Korean J-PARC Users, Grant No. NRF-2013K1A3A7A06056592).
\section*{Appendix}
The three-body phase space for the decay $ab\to123$ can be decomposed into the two-body one $ab\to1X\to23$ as follows:
\begin{eqnarray}
\label{eq:APP1}
\sigma_{ab\to123}&=&\mathcal{F}_{ab}\int d\Phi_3(ab\to123)|\mathcal{M}_{ab\to123}|^2
=\int d\Phi_2(ab\to1X)\frac{M^2_X}{2\pi}d\Phi_2(X\to23)|\mathcal{M}_{ab\to123}|^2
\cr
&=&\mathcal{F}_{ab}\int d\Phi_2(ab\to1X)\frac{dM^2_{23}}{2\pi}d\Phi_2(X\to23)\sum^\mathrm{spin}_X\frac{|\mathcal{M}_{ab\to1X}\mathcal{M}_{X\to23}|^2}{(M^2_{23}-M^2_X)^2+M^2_X\Gamma^2_X}
\cr
&\approx&\frac{\mathcal{F}_{ab}}{2M_X\Gamma_X}\int d\Phi_2(ab\to1X)d\Phi_2(X\to23)\sum^\mathrm{spin}_X|\mathcal{M}_{ab\to1X}\mathcal{M}_{X\to23}|^2
\cr
&=&\frac{\Gamma_{X\to23}}{\Gamma_X}\mathcal{F}_{ab}
\int d\Phi_2(ab\to1X)\sum^\mathrm{spin}_X|\mathcal{M}_{ab\to1X}|^2
=\frac{\Gamma_{X\to23}}{\Gamma_X}\sigma_{ab\to1X},
\end{eqnarray}
where $\mathcal{F}_{ab}$ stands for the flux factor for the initial state with $a$ and $b$. In deriving Eq.~(\ref{eq:APP1}), we have used the narrow-width approximation for the intermediate particle $X$: $\Gamma_X/M_X\ll1$, and taken into account that the invariant amplitudes $\mathcal{M}$ are insensitive to $M_{23}$. Considering the above decomposition, the invariant-mass distribution can be written by
\begin{eqnarray}
\label{eq:APP2}
\frac{d\sigma_{ab\to123}}{dM_{23}}&=&\mathcal{F}_{ab}\int d\Phi_2(ab\to1X)\frac{M_{23}}{\pi}d\Phi_2(X\to23)\sum^\mathrm{spin}_X\frac{|\mathcal{M}_{ab\to1X}\mathcal{M}_{X\to23}|^2}{(M^2_{23}-M^2_X)^2+M^2_X\Gamma^2_X}
\cr
&=&\mathcal{F}_{ab}\int d\Phi_2(ab\to1X)\frac{2M_XM_{23}}{2M_X\pi}d\Phi_2(X\to23)\sum^\mathrm{spin}_X\frac{|\mathcal{M}_{ab\to1X}\mathcal{M}_{X\to23}|^2}{(M^2_{23}-M^2_X)^2+M^2_X\Gamma^2_X}
\cr
&\approx&\mathcal{F}_{ab}\int d\Phi_2(ab\to1X)\frac{2M_XM_{23}}{\pi}\sum^\mathrm{spin}_X\frac{|\mathcal{M}_{ab\to1X}|^2\Gamma_{X\to23}}{(M^2_{23}-M^2_X)^2+M^2_X\Gamma^2_X}
\cr
&=&\frac{2M_XM_{23}}{\pi}\frac{\sigma_{ab\to1X}\,\Gamma_{X\to23}}{(M^2_{23}-M^2_X)^2+M^2_X\Gamma^2_X}.
\end{eqnarray}
Although there can be complicated interference effects due to other processes via a different intermediate particle $X'$, i.e. $ab\to 2X'\to 13$ for instance, for brevity, we ignored them here.  

\newpage

\newpage

\newpage

\newpage
\end{document}